\newcommand{\beq}{\begin{equation}}
\newcommand{\eeq}{\end{equation}}
\newcommand{\bea}{\begin{eqnarray}}
\newcommand{\eea}{\end{eqnarray}}
\newcommand{\bdm}{\begin{displaymath}}
\newcommand{\edm}{\end{displaymath}}
\newcommand{\te}{\theta}
\newcommand{\booeq}{ \mbox{ $    S_{cd}(\theta)=
              S_{ad}(\theta+\imath \overline{u}_{a\rev c}^{\rev b})
              S_{bd}(\theta-\imath \overline{u}_{\rev bc}^{\rev a}) $
                          }
                    }
\newcommand{\rev}[1]{\overline{#1}}

\newcommand{\esma}[2]{S_{#1}(\theta  #2 ) }
\newcommand{\inti}{ \int_{-\infty}^{+\infty} }
\newcommand{\pri}{ ^{\prime} }
\newcommand{\sumzi}{\sum_{s=0}^{\infty} }
\newcommand{\refe}[5]{{#1},\,{\em #2}\,{\bf #3}\,{(#4)}\,{#5}}
\newcommand{\sect}[2]{ \section{#1}\label{#2}  }
\newcommand{\sbl}[2] { [\frac{#1}{#2}] }

\newcommand{\square} {\framebox(2,2){} }
\documentstyle[12pt]{article}
\begin{document}
\titlepage
\title{
            From factorizable S-matrices to Conformal Invariance
\thanks{Lectures delivered at the  Jorge Andr\'e Swieca Summer School,
1990.}       }
\author{ Roland K\"{o}berle \thanks{ Supported in part by CNPq.} \\
 Instituto de F\'{\i}sica e Qu\'{\i}mica de S\~ao Carlos \\
Universidade de S\~ao Paulo\\
S\~ao Carlos 13560\\
BRASIL
       }
\maketitle
\newpage
\sect{Introduction}{s0}

Recently there has been spectacular progress in 1+1 dimensional
quantum field theory\cite{BPZ}. Although one may argue about
 whether this field
belongs to physics or mathematics, it is abundantly clear that a
successful non-perturbative attack on 4-dimensional field theories will
require a thorough understanding of the simpler counterpart in two dimensions.
This is about all we will say to justify  our living in so narrow an
environment, except perhaps to proclaim our believe that if someone finds
beautiful structures in mathematical physics, we may be sure that Nature will
make use of it.  And beautiful structures surely we will find. In the present
 lectures I'll try to make them as accessible as I can.

Trying to make progress in physics requires in general hard work. Yet this
can be very rewarding, if miracles do happen along the way - and in exactly
soluble models they do. After all some fine tuning of coupling constants
has taken place in order to guarantee solubility. One way to ensure this is
to impose factorizability on the $S$-matrix, i.e.  we require the $n$-particle
$S$-matrix to be a product of $2$-particle ones, consequently not allowing
particle production. This is guaranteed, if the model possesses an infinite
number of conservation laws \cite{consl,consl1}.

In section \ref{s1}, we will characterize our models by
 specifying their conservation
laws and then explain the {\em Bootstrap Principle} to construct factorized
$S$-matrices. This program is executed in section~\ref{s2} for some models.
In section~\ref{s3} the Thermodynamic Bethe Ansatz will be used
to connect these $S$-matrices to their ultra-violet limiting
conformally invariant field theory.
Section~\ref{s4} contains a short review of conformal invariance, so
that the unfamiliar reader doesn't have to go shopping somewhere
else, although in practice he may want to look up ref.\cite{CR}.
Finally in section~\ref{s5} we will study
perturbations of conformally invariant field theories, find out which
of them are candidates for exhibiting an infinite number of
conservation laws and thus make contact with section~\ref{s1}
\footnote{From
     a logical point of view,  this last section should come
     first, but it has to contain older, more technical material for
     which  excellent review articles exist \cite{CR,ZAMOT}.
     Therefore I have placed
     it near the end, hoping to delay as much as possible the loss of
     my audience. In this spirit I tried to exhibit - as we say in
     portuguese - 'o caminho das pedras'. }.
For the      benefit of the reader - hopefully -,  I have tried to
make these notes as
self-contained as possible,  at the risk of copying from too many
places already well known material.

\sect{1+1 dimensional factorizable $S$-matrix theory}{s1}

\subsection{Kinematics and scattering}

We will consider scattering of $n$ particles $A_{a}, \, a=1,2\ldots,n$, whose
masses are $m_{a}$. Their momenta satisfy the mass-shell condition
\beq
          p_{\mu}p^{\mu}=p\rev{p}=m^{2}  ,
\eeq
where the light-cone components of $p^{\mu}$ are \mbox{ $p=p^{0}+p^{1}$ } and
\mbox {$ \rev{p}=p^{0}-p^{1}$ }. The on-shell condition may be
conveniently parametrized introducing the rapidity $\theta\,  $:
\beq
          (p^{0}, p^{1})=(m\cosh\theta, m\sinh\theta).
\eeq

In order to ensure exact integrability we will assume that the field theory
giving rise to our $S$-matrix possess an infinite number of nontrivial,
commutative integrals of motion.

Although this is a well suited requirement in $d=2$ dimensions, for
 $d>2$ a
theorem by Coleman and Mandula \cite{CoMan} states that in  a non-trivial
Poincar\'e invariant field theory the most general invariance group is
a product of the Poincar\'e and an internal symmetry group. The proof
assumes some analyticity in energy and momentum transfer ( angle ).
This poses no problem in two dimensions,
since the scattering angle is anyhow only either $0$ or $\pi$.

These non-trivial, i.e.  other than energy-momentum,  conserved charges
$ Q_{\mu_1, \ldots,\mu_s } $ transform as $s$-th  order tensors under
the Lorentz group. In the light-cone representation we call them
 $ P_s, s=s_1,s_2,\ldots $, where the label $s$
indicates the spin of $P_s$. Actually in two dimension we have no rotation
group, so that {\em spin} refers to {\em Lorentz-spin}, specifying how $P_s$
behaves under a Lorentz transformations
\mbox {   $ L_{\alpha}: \theta\rightarrow \theta'=\theta+\alpha $ }. Thus
\beq
       P_s \rightarrow P'_s=e^{s\alpha}P_s.       \label{eq:Lorentz}
\eeq
For example the momentum $p$ has spin one : $p\rightarrow p'=e^{\alpha} p$
and the parity transformed $\rev{p} $ has spin minus one.
Therefore we have $P_1 = p $ and $P_{-1}=\rev{p} $.
Since parity
 relates the integrals of motion $P_{-s}$ to $P_{s}$, we will consider
only $s>0$ as we will deal only with parity conserving theories.

$P_s$ acts on one-particle states as
\beq
          P_s \mid A_a(p) > = \omega_s^a (p) \mid A_a(p) >.
\eeq
Since $P_s$ carries spin $s$ the Lorentz-transformation property
equ.(~\ref{eq:Lorentz}) requires $ \omega_s^a(p) $ to be of the form
\beq
              \omega_s^a(p) = \kappa _s^a p^s=\kappa_s^a(m_a)^se^{s\te},
\eeq
or
\beq
            P_s|A_a(p)>=\gamma_s^a|A_a(p)>,
             \hspace{3em} \gamma_s^a=\kappa_s^a(m_a)^s,
\label{eq:gamma}
\eeq
where $\kappa_s^a $ are real constants. For example $ \kappa_1^a=1 $.

$P_s$ are integrals of local densities. Therefore their action on
well separated multiparticle \em{ in } or \em { out  } states is
the sum of the one-particle contributions. Consequently  in a scattering
process
\beq
        p_{1}, \ldots,p_{n} \, \rightarrow \, p_{1}',\ldots,p_{m}'
\eeq
we have
\beq
      \sum_{i=1}^{n} (p_i)^s = \sum_{i=1}^{m} (p_i')^{s}.
\eeq
At least one non-trivial conservation law with $s>1$ implies in the
above equation $n=m$, \,\footnote{Except at most for a discrete set
                              of momenta, which we exclude appealing
                              to analyticity of the $S$-matrix \cite{consl1}. }
thus forbidding particle production and only
time-delays and exchange of quantum numbers are allowed \cite{consl}. Therefore
after an eventual relabeling, we have $p_i=p_{i}' $ and a product
of energy-momentum conserving   $\delta$-functions can be
 factored out of both terms of
the r.h.s.  of \makebox { $ S=I+\imath T $  }. When writing $S_{ab}(s)$
we will always assume these $\delta$-functions to have been factored
out.

The $n$-particle $S$-matrix is then a product of $n(n-1)/2$ two-particle
$S$-matrices. This decomposition can be effected
in several ways and all of them
must give the same result. The ensuing consistency conditions are called
Yang-Baxter factorization equations \cite{YB,ZZ}.

Let us order our particles along the line as
\mbox { $ p_1>p_2>\ldots>p_n $ }, so that {\em in}-states are arranged along
the space-axis according to decreasing  momenta and {\em out}-states the
opposite. Thus {\em in}-states are going to scatter, whereas particles in
{\em out}-states are running away from each other.
The action of the $S$-matrix on a
two-particle state is then defined as
\bdm
    \mid a(\theta_ 1 ), b(\theta_ 2 )>=
\edm
\beq
    S_{a\,  b}^{a' b'}(\te_1-\te_2)\mid a'(\te_2),b'(\te_1)>
 +\,S_{a \,b}^{b' a'}(\te_1-\te_2)\mid b'(\te_2),a'(\te_1)>.
\label{eq:2S}
\eeq
Note that the final states are ordered opposite to the initial states in
rapidity space, so that the $S$-matrix carries $in$-states into $out$-states.
This ordering introduces at most a phase, which is irrelevant for two-body
scattering, but when we write a many-body  amplitude as a product of
two-body amplitudes, this phase convention simplifies the factorization
equations.

For simplicity we assume
that in eq.(\ref{eq:2S}) the second { \em reflection } term vanishes and
only the first { \em transmission } term is present.
Unitarity requires then the $S$-matrix element
$S_{a b}^{a b} \equiv S_{a b} $ to be equal to
$ \exp \imath \delta(\theta) $ in the physical region.
The Yang-Baxter factorization
equations are now trivially satisfied for these diagonal $S$-matrices.
One way to ensure vanishing reflection is to require,  that  all particles
 $A_a$ have  different masses,  whenever $a \neq b$,
  i.e.  there is no degeneracy and no internal symmetry present,
( in particular particle and anti-particle
are identical $A_a=\overline{A}_a $). In this case no exchange
 of quantum numbers is possible. Or, if this is not the case, some other
mechanism, such as antiparticles
being bound-states of particles \cite{sw1}, particular conservation laws
 \cite{FAZA} etc. \  has to be invoked to ensure vanishing reflection.

In order to pin down the possible functional dependence of $S(\theta)$, we
have to discuss its analytical structure, which turns out to
be very simple.
In terms of the Mandelstam
variable $ s=(p_1+p_2)^2 \;\;S(s)$ has cuts along $ s\geq (m_1+m_2)^2 $ and
$ s \leq (m_1-m_2)^2 $, which are required by two-particle unitarity.
We suppose these to be the only cuts in the complex $s$-plane, since
there is no production and anomalous thresholds
turn into poles in $1+1$ dimensions \cite{anomthre}.
The cuts are eliminated by the uniformizing variable $\theta$ :
\beq
      s=s(\theta_1-\theta_2)=m_1^2+m_2^2+2m_1m_2\cosh\theta_{12},
\eeq
where $\theta_{12}=|\theta_1-\theta_2|  $. The mapping from the $s$- to
the $\theta$-plane is shown in figure~\ref{fig:s-plane}.

Since poles in the $s$-plane generate poles in the $\te$-plane,
 $\esma{ab}{}$ is a meromorphic function of $\te$.

Real analyticity in the $s$-plane ( Schwartz reflection principle ),
 ${\cal S}^{*}(s^{*})={\cal S}(s)$ ,
becomes ${\cal S}^{*}(-\te^{*})={\cal S}(\te)$ in the $\te$-plane.
In particular we see that the bound-state region
\mbox {$(m_1-m_2)^2 < s < (m_1+m_2)^2 $ } ,
where $S(s)$ is real, is mapped into the segment
\mbox {$ 0 <\Im m \theta < \pi $ } of the imaginary $\theta$-axis.
The unitarity condition
${\cal S}(s){\cal S}^{\dagger}(s)=
{\cal S}^{\dagger}(s){\cal S}(s)=I $ or
${\cal S}(\te){\cal S}^{\dagger}(\te)=
{\cal S}^{\dagger}(\te){\cal S}(\te)=I $,
then becomes for real $\te$
\beq
                  S_{ab}(\te )S_{ba}(-\te )=1.
\label{eq:unit}
\eeq
Parity invariance imposes
\beq
              S_{ab}(\te)=S_{ba}(\te)
\eeq
and charge-conjugation invariance implies
\beq
              S_{ab}(\theta)=S_{a\overline{b} }(\theta),
\eeq
where $ \rev{b} $ indicates the antiparticle of $b$. Finally
crossing-symmetry
\beq
    S_{a \rev{ b}}(s) = S_{\rev{ a} b}(s)= S_{ab}(2m_1^2+2m_2^2-s)
\label{eq:cross}
\eeq
becomes
\beq
      S_{\rev{a}b}(\te)=S_{a\rev{b}}(\theta)=S_{ab}(\imath\pi-\theta).
\eeq
It follows from equs. (\ref{eq:unit}) and (\ref{eq:cross}),
that $ S(\theta) $ is a $ 2 \pi \imath$- periodic function of
$\theta$.
\begin{figure}
\vspace{200pt}
\protect\label{fig:s-plane}
\begin{picture}(250,200)(-25,-160)
\thicklines
\put(200,0){\line(1,0){185}}
\put(200,-4){\line(1,0){185}}
\put(150,-2){\vector(1,0){245}}
\put(100,0){\line(-1,0){185}}
\put(100,-4){\line(-1,0){185}}
\put(200,-4){\line(0,1){4} }
\put(100,-4){\line(0,1){4} }
\put(400,-5){$\Re e\;\;s$}
\put(230,8){{\small right hand cut}}
\put(20,-14){{\small left hand cut}}
\put(200,-150){\line(0,1){100}}
\put(100,-150){\line(0,1){100}}
\put(150,-105){\vector(1,0){245}}
\put(400,-108){$\Im m\;\;\theta$}
\put(120,-95){{\small physical}}
\put(130,-119){{\small strip}}
\put(205,-102){$\theta=\imath\pi$}
\put(300,50){\framebox(60,15){s-plane}}
\put(300,-65){\framebox(60,15){$\theta$-plane}}
\multiput(120,-3)(15,0){5}{\square}
\multiput(120,-106)(15,0){5}{\square}
\end{picture}\\
\protect\caption{Mapping from $s$- to $\te$-plane,
showing the position of cuts and bound-state poles. }
\end{figure}
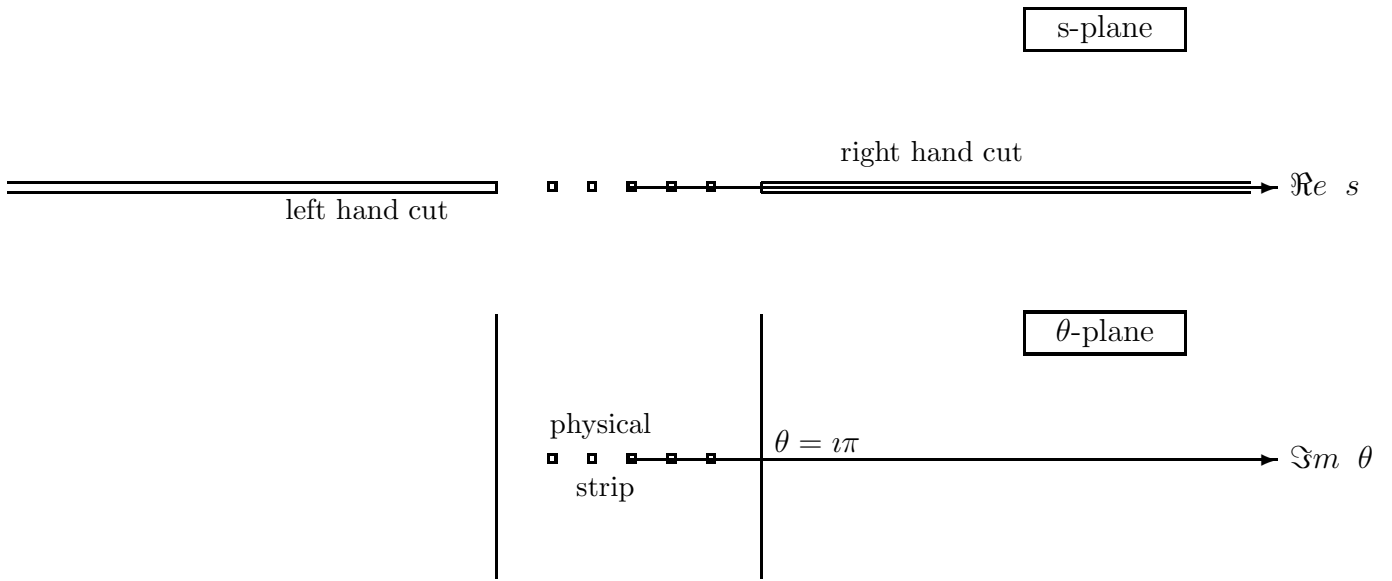
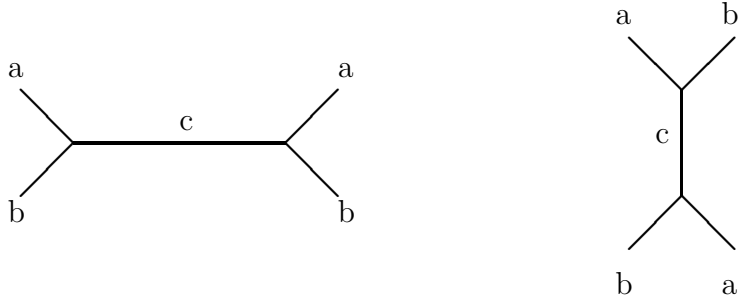
\begin{figure}
\vspace{60pt}
\begin{picture}(350,80)(-40,0)
\thicklines
\put(80,70){\line(1,0){80}}
\put(120,75){c}
\put(80,70){\line(-1,1){20}}
\put(55,95){a}
\put(80,70){\line(-1,-1){20}}
\put(55,40){b}
\put(160,70){\line(1,1){20}}
\put(180,95){a}
\put(160,70){\line(1,-1){20}}
\put(180,40){b}
\put(310,50){\line(0,1){40}}
\put(300,70){c}
\put(310,50){\line(1,-1){20}}
\put(325,13){a}
\put(310,50){\line(-1,-1){20}}
\put(285,13){b}
\put(310,90){\line(1,1){20}}
\put(325,115){b}
\put(310,90){\line(-1,1){20}}
\put(285,115){a}
\end{picture}\\
\protect\caption{Diagrams responsible for poles
in the direct and crossed channel. }
\protect\label{fig:bsdiagr}
\end{figure}
We now exhibit the general solution $S_{ab}(\theta)$ of these
equations. Since we know that the scattering amplitudes are bounded
functions of the momenta, one can show \cite{mitra} that this implies
that any meromorphic, real analytic, $ 2 \pi \imath$-periodic function
$ f(\theta) $ satisfying $ f(\theta)f(-\theta)=1 $,
can be represented as
\beq
      f(\theta)=\prod_{\alpha\in{\cal A}} f_{\alpha}(\theta),
\hspace{2em} with\hspace{1em}
 f_\alpha(\theta)=\frac{\sinh[1/2(\theta+\imath\pi\alpha)]}
                       {\sinh[1/2(\theta-\imath\pi\alpha)]},
\eeq
where ${\cal A}$ is a set of complex numbers invariant under complex
conjugation. If there are no unstable particles present,  then all poles
of $f(\theta)$ occur on the imaginary $\theta$-axis, implying $\alpha$
to be real and we may restrict it to  $-1<\alpha\leq+1 $. In this case
$ f(\theta) $ has a simple pole of residue
$ 2\imath\sin\alpha\pi $ at
$ \theta=\imath\alpha\pi $ and a simple zero at
 $\theta=-\imath\alpha\pi$.
Besides this we note the following useful properties:
\bdm
          f_{\alpha}(\theta)  =  f_{\alpha+2}(\theta)=f_{-\alpha}(-\theta)
\edm
\bdm
          f_{\alpha}(\theta)f_{-\alpha}(\theta) =
           f_{\alpha}(\theta)f_{\alpha}(-\theta) =1
\edm
\beq
         f_{\alpha}(\imath\pi -\theta)=- f_{1-\alpha}(\theta)
\label{eq:fff}
\eeq
\bdm      f_{\alpha}(\theta-\imath\pi\beta)
          f_{\alpha}(\theta+\imath\pi\beta)
           =f_{\alpha+\beta}(\theta)f_{\alpha-\beta}(\theta)
\edm
\bdm
         f_0(\theta)\equiv-f_1(\theta)\equiv 1\;.
\edm
If at least one of the particles $A_a$ or $A_b$ is self-conjugate, then
crossing symmetry
\beq
                S_{ab}(\theta)=S_{ab}(\imath\pi-\theta)
\eeq
implies,  that up to a sign $S_{ab}(\theta)$ must be of the form
\beq
   F_{\alpha}(\theta)=f_{\alpha}(\theta)
                      f_{\alpha}(\imath\pi-\theta)=
   \frac{sinh\theta+\imath\sin\alpha\pi}
        {sinh\theta-\imath\sin\alpha\pi}=
   \frac{\tanh[1/2(\theta+\imath\alpha\pi)]}
        {\tanh[1/2(\theta-\imath\alpha\pi)]}.
\eeq
The functions $F_{\alpha}(\theta)$ satisfy:
\bdm
     F _{\alpha}(\theta)=F_{\alpha+2}(\theta)=F_{1-\alpha}(\theta)
     =F_{-\alpha}(\theta)
\edm
\beq
    F_{\alpha}(\theta)F_{-\alpha}(\theta) \;=\;1
\label{eq:F}
\eeq
\bdm
     F_{\alpha}(\theta+\imath\pi\beta)
     F_{\alpha}(\theta-\imath\pi\beta)=
     F_{\alpha+\beta}(\theta)F_{\alpha-\beta}(\theta)
\edm
\bdm
       F_0(\theta) \equiv 1\;.
\edm
For $ 0<\alpha<1/2 $ , $ F_{\alpha}(\theta) $  has simple poles at
$\imath\alpha \pi$ and $\imath (1-\alpha)\pi$ with residues
 $2\imath\tan\;\alpha\pi$
and $-2\imath\tan\;\alpha\pi$, respectively, as well as zeroes at
$-\imath\alpha\pi$ and $-\imath (1-\alpha)\pi$.
$F_{\frac{1}{2}}(\theta)$ has a
double pole at $\imath \pi /2 $ and a double zero at $\;-\imath\pi/2$.

Since bound-states will be all important in setting up the bootstrap
principle,  let us have a fast look at them,  both in the $s$- and the
$\theta$-plane. In the $s$-plane $S_{ab}(s)$ has bound-state poles
for \makebox{ $ 0<s=m_c^2<(m_a+m_b)^2 $}, whose residues are positive due to
unitarity :
\beq
          S_{ab}(s) \sim  (f_{ab}^{\rev{c}})^2/(s-m_c^2).
\eeq
When we consider $S_{ab}(\theta)$, we suppose always that the
 $\delta$-functions of the form \makebox{ $\prod\delta(\theta_i-\theta_j) $}
have been factored out. Therefore in going from $S(s)$ to
$S(\theta)$,  we have to take into account the
 Jacobian

\beq
         \frac{ \partial(p_a, p_b)}{\partial(\theta_a, \theta_b)}
           = m_a m_b\sinh(\theta_{ab}).
\eeq
Therefore near the position of the pole in the $\theta$-plane given by
\beq
     m_c^2= s(\theta=iu_{ab}^{c})=
      m_a^2+m_b^2+2m_am_b\cos u_{ab}^{c} ,
\label{eq:mc}
\eeq
i.e.  near $\theta=\imath u_{ab}^c$, $S_{ab}(\theta) $ behaves as
\beq
         S_{ab}(\theta) \sim
         \imath R_{ab}^c/(\theta-\imath u_{ab}^c),
\eeq
where $R_{ba}^c=f_{ab}^{\rev{c}}f_{ab}^{\rev {c}}$.
{}From crossing-symmetry we see,
that there is also a pole with negative residue $ - R_{ab}^c $ at
$\theta=\imath\rev{u}_{ab}^c=\imath(\pi-u_{ab}^c) $. This behavior is
caused by the diagrams displayed in figure~\ref{fig:bsdiagr}, which
shows, that $f_{abc} \equiv f_{ab}^{\rev {c}}$ is a completely
symmetric function of its indices. From the law of cosines applied to
the triangle with sides $m_a, m_b,m_c$, we obtain from equ.(\ref{eq:mc})
\beq
       u_{ab}^c+u_{\rev c a}^{\rev b}a+u_{b \rev c}^{\rev a}=2\pi.
\label{eq:3u}
\eeq

Up to now we have listed the properties of $\esma{ab}{}$ following
from general principles. These are of course insufficient to
construct the two-particle $S$-matrices.
Since the factorization equations are void for diagonal $S$-matrices,
we need another input.
This is furnished by the {\em Bootstrap Principle }. Suppose our theory
contains the set of particles $ A_1, \ldots, A_n $. Then we require that the
bound-states of $S_{ab} (\theta) $ with positive residue represent one of
the particles $A_a$ and vice-versa - any particle $A_a$ occurs as a
bound-state in some $S$-matrix element.

Suppose a particle $ A_c $ occurs as a bound-state in the scattering of
$A_a A_b$ at $\theta_{ab}=\imath u_{ab}^c$. Then scattering $A_c$ with $A_d$
must give the same result as the three-particle scattering
\beq
 S_{abd}(\theta_a, \theta_b,\theta_c)=
 S_{ab}(\theta_{ab})S_{ad}(\theta_{ad}) S_{bd}(\theta_{bd} )
\eeq
at the relative rapidity $\theta_{ab}=\imath u_{ab}^{c}$ after
dividing the $S$-matrix  by the residue $\imath (f_{abc})^2$.

The kinematics goes as follows. Using equ.(\ref{eq:3u}), we can write
$ \te_{ab}=\imath u_{ab}^{c}=\imath (2 \pi - u_{\rev ca}^{\rev b}
 - u_{b\rev c}^{\rev a} )=
\imath ( \rev{u}_{\rev ca}^{\rev b} + \rev{u}_{b\rev c}^{\rev a} )$.
 This suggests to apportion rapidities as :
\bdm
 \te_a=\rev\te+\imath\rev{u}_{\rev ca}^{\rev b},\hspace{1cm}
  \te_b=\rev\te-\imath\rev{u}_{b\rev c}^{\rev a},
\edm
where $\rev\te$ is the center of mass rapidity of the $ab$-system.
Hence we get
\bdm
    \te_{ad}=\rev\te-\te_d + \imath\rev{u}_{\rev ca}^{\rev b}=
     \te+\imath\rev{u}_{\rev ca}^{\rev b}
\edm
\bdm
    \te_{bd}=\rev\te-\te_d - \imath\rev{u}_{b\rev c}^{\rev a}=
    \te+\imath\rev{u}_{b\rev c}^{\rev a}.
\edm

Thus the simplest equation
compatible with the above requirement is the {\em bootstrap
equation }
\beq
\label{eq:bs}
\booeq.
\eeq
It is this equation we will use as a consistency requirement to
construct $S$-matrices in the next section.

The bootstrap equation generates a powerful constraint on the possible
conservation laws \cite{ZAMOT}. Namely,
if we continue the two-particle state
$  \mid A_a(\theta_a) A_b(\theta_b) > $ to
 $\theta_{ab}=\imath u_{ab}^c$ it
will be dominated by $\mid A_c>$, i.e.
\beq
     \lim_{\epsilon\rightarrow0} \epsilon \mid
     A_a(\theta+\imath\rev{u}_{a\rev c}^{\rev b}-\epsilon/2)
     A_b(\theta-\imath\rev{u}_{b\rev c}^{\rev a}+\epsilon/2)>
   = f_{ab}^{\overline{c} } \mid A_c(\theta) >.
\eeq
This has to be compatible with our conservation laws
$ [S,P_s]=0 $. If we set $\te_a=\te + \imath \rev u_{a\rev c}^{\rev b}
 - \epsilon/2$ and $ \te_b=\te - \imath \rev u_{b\rev c}^{\rev a}
+ \epsilon/2$, then the equation
\beq
     P_s \mid A_a(\te_a) A_b(\te_b) >_{in} =
   (\gamma_s^a e^{s\te_a} + \gamma_s^b e^{ s\te_b })
   \mid A_a(\te_a) A_b(\te_b) >_{in},
\eeq
when continued analytically to the pole,  yields
\bdm
  P_sS|A_a(\te_a)A_b(\te_b)>_{\em in} =
  P_s \frac{f_{abc}}{\epsilon}|A_c(\te)> =
  \frac{f_{abc}}{\epsilon}\gamma_s^c e^{s\te}|A_c(\te)>=
\edm
\bdm
    SP_s|A_a(\te_a)A_b(\te_b)>_{\em in} =
    S[\gamma_s^a e^{s\te_a}+\gamma_s^b e^{s\te_b}]
    |A_a(\te_a)A_b(\te_b)>_{\em in}=
\edm
\beq
   [\gamma_s^a e^{s\te_a}+\gamma_s^b e^{s\te_b}]
    \frac{f_{abc}}{\epsilon}|A_c(\te)>.
\eeq
Therefore we get
\beq
      \gamma_s^a e^{ -\imath s\rev{u}_{a\rev c}^{\rev b}}
     +\gamma_s^b e^{ +\imath s\rev{u}_{\rev bc}^{\rev a}} =
      \gamma_s^c,
\label{eq:bc}
\eeq
valid for all $f_{abc}\neq 0$ .
If all angles $ u_{ab}^c $ are known, this equation can in general be
solved for the $\gamma_a^s=\kappa_a^s (m_a)^s,\ldots $ {\em only at certain
values} of $s$, determining in this way the possible conservation laws.
\sect{Factorized $S$-matrices}{s2}

Let us use the prescription of the preceeding section and construct
some interesting $S$-matrices\cite{TKEM}. Of course we still have to feed in
information characterizing some particular model,  such as some
minimum particle content, symmetry properties, simplicity etc.
In sections \ref{s3} to \ref{s5} we will link up with the
underlying field theory.

\subsection{The $S$-matrix of the Lee-Yang edge singularity}

Consider the simplest possible model, where we have only one self-conjugate
particle $A_1=A_{\rev 1}$,
which according to the bootstrap principle can be considered as an $A_1 A_1$
bound-state of itself. This means that $ f_{111}\neq0 $ and this will be
referred
to as $A_1$ having the \, $\phi^3\;${\em  property} for obvious reasons.
Thus setting $a=b=c=1$ in equ.(\ref{eq:bc}) and using
\makebox{$  u_{11}^1 = \frac{2\pi}{3} $} from equ.(\ref{eq:3u}), we get
\beq
         e^{ -\imath\pi s/3} + e^{ +\imath\pi s/3} = 1.
\eeq
Rewriting this as $\cos(\pi s/3)=1/2$, we see that this equation is satisfied
only for $ s=6n \pm 1 $, i.e.  $s=1, 5, 7, 11, 13, 17, 19, 23, 25, \ldots$.

Since we have a pole at $\theta=2\pi \imath /3$ the
 {\em minimal } solution to the
equ.(\ref{eq:bs}) is
\beq
        S_{11}(\theta)=F_{\frac{2}{3}}(\theta).
\label{eq:sLY}
\eeq
The residue of this amplitude at $\theta=2\pi \imath /3$ equals $+2i\sqrt{3}$.
Thus,
although the $S-$matrix satisfies the correct unitarity equation, the theory
as a whole is not unitary, since the residue has got the wrong sign. The
resolution of this problem goes noting that for the unitarity equation
${\cal SS}^{\dagger}=1  $ to hold it is sufficient, that the {\em in-}
and {\em out-}states form a complete set. In fact
from $ S|{\em in}>=|{\em out}> $ and $<{\em in}|S^{\dagger}=<{\em out}|
$ we get $<{\em in}|S^{\dagger}S|{\em in}>=<{\em out}|{\em out}>=1 $.
{}From the completeness of the {\em in}-states we conclude that
{\cal SS}$^{\dagger}=1$ and similarly for {\cal S}$^{\dagger}${\cal S}$=I $.
The hamiltonian doesn't have
to be hermitian \cite{cm-ly}.

We   conclude,  that a non-trivial unitary theory
with a factorizing $S-$matrix has to contain at least two particles.

This model is only the first one of a whole series of non-unitary
models\cite{CHIC}, as we will show after discussing the $Z(N)$-models.

As we will see, the $S$-matrix equ.(\ref{eq:sLY}) belongs to the
non-unitary, conformal field theory describing the Lee-Yang edge
singularity\cite{lee-yang}, deformed by the only relevant operator it
contains.

\subsection{The Ising model in a magnetic field}

Since one particle is not sufficient to produce a unitary theory, let
us go one step further and consider an extra particle $A_2$\cite{ZAMOis}.
 As above
let us assume that $A_1$ has the $\phi^3$ property and also couples
to the $A_2$ particle, i.e.  assume that $f_{111}, f_{112} $ and
$ f_{221} $ do not vanish. We will furthermore assume all particles to be
self-conjugate.

Using now equ.(\ref{eq:bc}) with $ a=b=1;c=2 $ we get
\begin{equation}
     \kappa_s^1 m_1^s e^{-\imath s\rev{u}_{12}^1}
    +\kappa_s^1 m_1^s e^{+\imath s\rev{u}_{12}^1} =
     \kappa_s^2 m_2^s.
\end{equation}
 Repeating the same for $ a=b=2;c=1 $ results in
\beq
     x_1^s + x_1^{-s} = \left( \frac{m_2}{m_1} \right)^s
                               \frac{\kappa_s^2}{\kappa_s^1},
     \hspace{4em}
     x_2^s + x_2^{-s} = \left( \frac{m_1}{m_2} \right)^s
                               \frac{\kappa_s^1}{\kappa_s^2},
\eeq
where
\beq
     x_1=e^{\imath \rev{u}_{12}^1} \:;\:
     x_2=e^{\imath \rev{u}_{21}^2}.
\eeq
Eliminating the r.h.s.  of these two equations,  we finally obtain
\beq
       ( x_1^s+x_1^{-s} ) ( x_2^s+x_2^{-s} )= 1.
\label{eq:x12s}
\eeq
This is a very much overdetermined system of equations, if we
have non-trivial conservation laws with $s>1$. But, as we mentioned,
miracles do happen and equs.(\ref{eq:x12s}) do have a solution for $s$
belonging to a subset of integers. For $s=1$, we get :
\beq
    x_1=\exp(\pi i/5),\hspace{1cm}  x_2=\exp(2\pi i/5).
\eeq
This solution can be obtained making the ansatz $x_2=(x_1)^r$.
$r=1$ gives $m_1=m_2$, which we don't want, whereas the above solution
corresponds to $r=2$. For $s=1$ we therefore obtain the {\em golden} mass ratio
\beq
        \frac{m_2}{m_1} = 2\cos\frac{\pi}{5}=\frac{\sqrt{5}+1}{2}=
                      1.6180339\ldots.
\eeq
With this information, we can write equ.(\ref{eq:x12s}) as
\beq
   \cos(\frac{\pi}{5})\cos(\frac{2\pi}{5})=
   \cos(\frac{s\pi}{5})\cos(\frac{2s\pi}{5}).
\eeq
This equation is satisfied only for $s\neq 0 (mod\, 5)$.
Recalling that $A_1$ has the $\phi^3$ property, the allowed values for
$s$ are now
\beq
             s=1, 7, 11, 13, 17, 19, 23, 29, 31, \ldots.
\label {eq:Is}
\eeq
We may now start to use the bootstrap equation (\ref{eq:bs}) to crank out
the possible $S$-matrices of this model.
Start with $S_{11}$. It is supposed to have poles at
$\theta=\frac{2\pi \imath}{3}$
and $\theta=\frac{2\pi \imath}{5}$ with positive residues corresponding to
the particles $A_1$ and $A_2$ in the direct channel and corresponding
poles with negative residues in the crossed channel. Using our functions
$ F_{\alpha}(\theta) $ (which we abbreviate as $ [\alpha] $ ) as building
blocks,  we choose the following $0$'th order trial for $S_{11}$ :
\beq
          S_{11}^{(0)}=\sbl{2}{3}\;\sbl{2}{5}.
\eeq
$S_{11}^{(0)}$ has to satisfy the bootstrap equation (\ref{eq:bs}),
whose r.h.s.  becomes :
\beq
        \sbl{1}{3}\;[1]\;\sbl{1}{15}\;\sbl{11}{15}.
\eeq
Here we used equ.(\ref{eq:F}) to maneuver the displaced $\theta$-dependence
into the indices. Due to the presence of $[1/15]$ and $[11/15]$ , this cannot
be equal to the l.h.s.  of the bootstrap equation. Therefore we modify
$S_{11}^{(0)} $ to what will be our final version:
\beq
        S_{11}(\theta)=\sbl{2}{3}\;\sbl{2}{5}\;\sbl{1}{15}.
\eeq
It is now easily seen, that the bootstrap equation is satisfied at the
expense of having introduced new poles into $S_{11}$,  which depending
on the sign of the residues,  will correspond to new particles. As a matter
of fact the new pole at $\theta=\imath u_{11}^3=\pi \imath /15$
 represents a new
particle $A_3$.

Note that up to now we have from $A_1A_1\rightarrow A_1$,
$A_1A_1 \rightarrow A_2,A_1 A_1 \rightarrow A_3 $ the following results:
\bdm
                u_{11}^1=\frac{2\pi}{3}, \hspace{3em}
                u_{11}^2=\frac{2\pi}{5}, \hspace{3em}
                u_{11}^3=\frac{\pi}{15},
\edm
\beq
                u_{12}^1=\frac{4\pi}{5},  \hspace{3em}
                u_{21}^2=\frac{2\pi}{5} .
\eeq
Thus from equ.(\ref{eq:mc}) we obtain
\beq
       \frac{m_3}{m_1} = 2 \cos \left( \frac{\pi}{30} \right) =
                       1.9890437\ldots.
\eeq
Now we have to continue turning the crank, hoping that eventually
no new particles will be necessary in order to satisfy the bootstrap
equation equ.(\ref{eq:bs}). In the meanwhile,  we use it with
$ a=b=d=1, c=2$ :
\beq
  \esma{12}{} =   \esma{11}{-\pi \imath/5}
                  \esma{11}{+\pi \imath/5}   .
\eeq
This yields
\beq
     \esma{12}{} = \sbl{4}{5}\;\sbl{3}{5}\;\sbl{4}{15}\;\sbl{7}{15}.
\eeq
We have a new pole with positive residue at $\theta=
\imath u_{12}^4=4\pi \imath/15 $,
corresponding to a new particle $A_4$ with mass
\beq
     \frac{m_4}{m_1} = 4\cos\left( \frac{ \pi}{5}\right)
                        \cos\left( \frac{7\pi}{30}\right)
                     = 2.4048671\ldots.
\eeq
The amplitude $S_{22}$ can now be obtained from equ.(\ref{eq:bs}) with
$a=b=1, c=d=2$ :
\beq
    \esma{22}{} =  \esma{12}{-\pi \imath /5}
                   \esma{12}{+\pi \imath /5}.
\eeq
This yields
\beq
      \esma{22}{}= \esma{11}{}\esma{12}{}.
\eeq

$\esma{22}{}$ exhibits two positive-residue poles at
$ \theta= \imath u_{22}^5=4\pi \imath /14$ and
$ \theta=\imath  u_{22}^6=\pi \imath /15 $~,
representing new particles $A_5$ and $A_6$ with masses:
\bea
     \frac{m_5}{m_1}&=& 4\cos\left( \frac{ \pi}{5}\right)
                         \cos\left( \frac{2\pi}{15}\right) =
                     2.957295\ldots
\nonumber\\
     \frac{m_6}{m_1}&=& 4\cos\left( \frac{ \pi}{5}\right)
                         \cos\left( \frac{\pi}{30}\right) =
                         3.2183404\ldots.
\eea

We notice, that double poles occurring in, for example $ \esma{22}{} $,
  correspond to anomalous thresholds \cite{anomthre}.

Let us go on and construct $\esma{13}{}=\esma{31}{} $, putting
$a=b=d=1$ and $c=3 $ in equ.(\ref{eq:bc}). We need $u_{13}^1 $ ,which we
lift from equ.(\ref{eq:3u}), using $  u_{11}^3=\pi /15 $ and
$   u_{ab}^c=u_{ba}^c $. We obtain
\beq
   \esma{13}{} = \esma{11}{+\imath \pi /30}
                 \esma{11}{-\imath \pi /30},
\eeq
which yields
\bdm
  \esma{13}{}= \sbl{21}{30}\;\sbl{19}{30}\;\sbl{13}{30}\;\sbl{11}{30}
               \;\sbl{1}{10}\;\sbl{1}{30}
\edm
\beq
      \sbl{1}{10}\;\sbl{1}{30}\;
       \left (\sbl{11}{30} \right)^2\;\sbl{13}{30}\;\sbl{21}{30}.
\eeq
In a similar way we get :
\bea
   \esma{23}{} & = & \sbl{1}{6}\;\sbl{19}{30}\;\sbl{3}{10}
                    \;\left(\sbl{7}{30}\;\sbl{13}{30}\;\sbl{1}{2}\right)^2
                           \nonumber \\
   \esma{33}{} & = & \sbl{2}{3}\;\sbl{2}{15}\;\sbl{2}{10}\;
                      \esma{11}{}\esma{22}{}  \\
   \esma{44}{} & = & \sbl{2}{15}\;\sbl{7}{15}\;\sbl{2}{3}^2\;
                      \esma{12}{}\esma{22}{} .
                            \nonumber
\eea
We encounter two new particles : $A_7$ as a pole at
$\theta=2\pi \imath /15 $
in $ \esma{33}{} $ and $A_8$ as a pole at $\theta=\pi \imath /15 $
in $ \esma{44}{} $ , whose masses are :
\bea
     \frac{m_7}{m_1}&=& 8\cos^2\left( \frac{ \pi}{5}\right)
                         \cos  \left( \frac{7\pi}{30}\right) =
                     3.891156\ldots    \nonumber \\
     \frac{m_8}{m_1}&=& 8\cos^2\left( \frac{ \pi}{5}\right)
                         \cos  \left( \frac{2\pi}{15}\right) =
                         4.783386\ldots.
\eea

At this point we may rest, since to our deep satisfaction we encounter
{\em no new poles } and the bootstrap program closes with 8 of them.
As we shall see these $S$-matrices describe the massive field theory,
obtained by perturbing the critical $T=T_c$ zero-field Ising model by a
magnetic field.
We also  notice that an underlying
structure pertaining to the exceptional  Lie algebra $E_8$ raises it's
delightful countenance :
the conservation laws equ.(\ref{eq:Is}) are labeled by spins,
which are exactly the {\em exponents} of the Lie algebra $E_8$,
repeated modulo the {\em Coxeter } number of $E_8$ \cite{FAZA}.
This is not completely unexpected, since the conformal field theory
describing the critical Ising model can be obtained via the
{\em coset} construction\cite{GKO} $(E_8)_1\bigotimes(E_8)_1/(E_8)_2$,
where the subscript denotes the {\em level} of the Kac-Moody algebra.

\subsection{The $Z(N)$-models}

This set of interesting models are the simplest generalization of
the Ising model, which corresponds to $N=2$.
Their symmetry is actually $Z(2)\times Z(N)$, where the extra $Z(2)$
factor stands for charge conjugation.
They exhibit a very rich and interesting phase
diagram\cite{ALKO}, which includes special multicritical points,
where exact solutions can be obtained, even off criticality.
These points then become relevant for the $S$-matrix game.

On a lattice these models are defined by spins $\sigma_i$
living on each site $i$, which satisfy $ (\sigma_i)^N=1 $ or
equivalently $\sigma_i^{\dagger}=(\sigma_i)^{N-1} $. In the field-theoretic
context this is translated into the  property, that anti-particles are
bound-states of $N-1$ particles \cite{KOSW}. This property requires the
reflection amplitude to vanish, since otherwise this would correspond to
non-vanishing production in the crossed channel.

For the scattering of  {\em fundamental } particle, we
have then only two amplitudes
\beq
       <p_2,p_1|S|p_1,p_2>=\esma{11}{_{12}}
\eeq
\bdm
       <\rev {p}_2,p_1|S|p_1,\rev{p}_2>=\esma{1\rev 1}{_{12}}.
\edm
Unitarity and crossing imply
\beq
    u(\te)u(-\te)=t(\te)t(-\te)=1
\eeq
\bdm
     u(\te)=t(\imath \pi - \te).
\edm
We now make a minimality assumption and introduce a pole,
 corresponding to a two-particle bound-state at
$\te_{12}=\imath u_{11}^2=2 \pi \imath/N$  in $u(\te)$, the following
mass spectrum is generated \cite{STW}:
\beq
      m_a=m\frac{\sin(\pi a/N)}{\sin(\pi/N)},\hspace{3em}
           a=1,\ldots,N-1.
\label{eq:ZNM}
\eeq
Here we have $m_{N-1}=m$ or more generally,
 in agreement with charge-conjugation invariance, $m_{N-a}=m_a$.

In the present model, as opposed to the Ising case with $h\neq 0,T=T_c$,
we know the mass spectrum and therefore don't have to go through all
the motions to find $u_{ab}^c$.
We immediately obtain the
{\em minimal} $S$-matrix
of the 'fundamental' particle \cite{KOSW}:
\beq
       \esma{11}{}=f_{\frac{2}{N}}(\te).
\label{eq:z1}
\eeq

For $N=3$ just apply the equ.(\ref{eq:bs}). For larger $N$
one has to fuse $N-1$ particles to check, that anti-particles are
bound-states of $N-1$ particles \cite{KOSW}, which means that
the following identity must hold :
\beq
 \prod_n u(\te+n\imath \pi/N)=t(\te)=u(\imath\pi-\te),
\eeq
where
\bdm
     n=\left\{ \begin{array}{ll}
             \,\,\,\,\pm1,\pm3,\ldots,\pm(N-2) for N odd\\
             0,\pm2,\pm4,\ldots,\pm(N-2) for N even.
                \end{array}
                \right.
\edm
For $N=2$ this gives $S_{11}(\te)=-1$,  as it should for a free bosonic
theory.

The complete two-particle $S$-matrix is obtained via the Bootstrap
Principle as
\beq
   \esma{ab}{}=f_{\frac{|a-b|}{N}}(\te)\left[
                       \prod_{k=1}^{min(a, b)-1} f_{\frac{|a-b|+2k}{N}}(\te)
                               \right]^2
               f_{\frac{(a+b)}{N}}(\te),
\label{eq:z2}
\eeq
where $a, b=1, 2, \ldots, N-1$.
The simple poles of $\esma{ab}{}$, which would violate the $Z(N)$-symmetry,
 if interpreted as particles, are doomed to exhibit residues with the
wrong sign.

We can now check for which values of $s$ we encounter conservation laws.
The mass spectrum together with equ.(\ref{eq:mc}) yields
\beq
    u_{ab}^c=\frac{\pi (a+b)}{N},\;\;\;c=a+b\,(mod\, N).
\eeq
Using  this information in equ. (\ref{eq:bc}),  we arrive at - recall
$\rev a=N-a$ ! -
\beq
    \gamma_s^a e^{-\imath s \pi b/N}+ \gamma_s^b e^{+\imath s \pi a/N}=
    \gamma_s^{a+b}.
\eeq
Since the $\gamma_a^s$ are real, the imaginary part of the l.h.s.
has to vanish, yielding the following condition:
\beq
           \gamma_s^b \sin(s\pi b/N)=\gamma_s^a \sin(s\pi b/N)
\label{eq:kbs}
\eeq
This therefore better be an $a,b$-independent constant,which
we normalize  to unity and get:
\beq
      \gamma_s^a=\sin(s\pi a/N).
\label{eq:gama}
\eeq
The real part gives:
\beq
   \gamma_s^a\cos(s\pi b/N)+
   \gamma_s^b \cos(s\pi a/N)=\gamma_s^{a+b}\sin[\pi(a+b)/N]
\eeq
and this  is identically satisfied for $\gamma_s^a$ from equ.(\ref{eq:kbs}).
All this is of course consistent only for
$\sin(\pi sa/N)\neq0$, or
\beq
      s \neq 0 (mod N)
\eeq
and we get an infinite number of conservation laws for all these values of $s$.
Let us expose, that this set reveals an underlying structure corresponding to
the group $A_{N-1}=SU(N)$.

The relevant properties of {\em simple Lie algebras} can be encoded in
 {\em Dynkin} diagrams\cite{LIE}. For $A_{N-1}$ or $SU(N)$ this looks
like
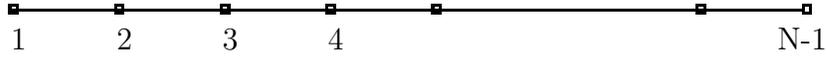
\newline
\begin{picture}(400,27)(-55,0)
\thicklines
\put(0,15){\line(1,0){300}}
\multiput(0,14)(40,0){5}{\square}
\put(260,14){\square}
\put(300,14){\square}
\put(0,0){1}
\put(40,0){2}
\put(80,0){3}
\put(120,0){4}
\put(290,0){N-1}
\end{picture}\\
\newline
The incidence matrix $I_{ab}$ of this graph is the $(N-1)*(N-1)$ matrix

$ \hspace{6em} I_{ab}=$  number of lines between nodes a and b
\beq
    I=\left(\begin{array}{ccccc}
                     0 & 1 & 0 & \ldots & 0 \\
                     1 & 0 & 1 & \ldots & 0  \\
                     0 & 1 & 0 & \ldots & 0   \\
               \vdots  & \vdots & \vdots & \vdots & \vdots \\
                     0 & 0 & 0 &    1   & 0
                   \end{array}\right )
\eeq

The eigenvalues of $A$ are
\beq
    i^{(s)}=2\cos\frac{s\pi}{N}=\frac{\sin\frac{2s\pi}{N}}{\sin\frac{s\pi}{N}}
\eeq
for $s=1,2,\ldots,N-1$. The corresponding normalized eigenvectors are
\beq
        \vec\Gamma_s=\Gamma_s^{(a)}=\sqrt{\frac{2}{N}}\sin\frac{sa\pi}{N}.
\eeq
The matrix $I$ has non-negative entries and one defines the
{\em Perron-Frobenius } vector as the unique eigenvector, all of
 whose components can be chosen to be positive. The corresponding eigenvalue
is not smaller in magnitude than any other eigenvalue\cite{LIE}.
In our case it is the vector $\vec\Gamma_1$.

At this point, we take a rest and compare our $Z(N)$ mass formula
equ.(\ref{eq:ZNM}) and our result for $\gamma_s^a$ equ.(\ref{eq:gama})
with the preceeding algebraic constructs.
We immediately notice:

 i) $\Gamma_1^{(a)}=
\sqrt{\frac{2}{N}}\sin\frac{a\pi}{N}$ and therefore {\em the components
of the Perron-Frobenius vector} give - up to normalization - the mass
spectrum.

 ii) $\Gamma_s^{(a)}=\gamma_s^a $ or $\vec\Gamma_s=\vec\gamma_s$, where
$\vec\gamma_s=\gamma_s^1,\ldots,\gamma_s^{N-1}$. The eigenvalues of
the conserved charges $P_s$ are thence also given by group theory.

If we replace in the above formulas the number $N$ by the {\em Coxeter}
number $ h^{\cal A} $ of the group ${\cal A}$, then these two results are
valid for all the simple-laced groups ${\cal A,D,E}$. ${\cal E}_8$ corresponds
to the Ising model in a magnetic field and all the other cases have also been
identified\cite{FAZA,TODA}.

We now want to reap some rewards by extending our solutions a bit.
 All the $S$-matrices obtained in this section
are {\em mimimal} in the sense, that they satisfy all the imposed constraints
with a minimum number of poles. However there are models with the
same symmetries, but which contain e.g. a free parameter. A very prominent
set are the {\em Toda } field theories\cite{TODA1},
 which are exactly integrable and
contain a coupling parameter $\beta$, on which the $S$-matrices must
depend. As first shown in ref.\cite{AFZ}, one can find solutions of
the relevant equations, which are products of two factors : one
equals the minimum $S_{ab}(\te)$-matrix and
 the other $Z_{ab}(\te)$ contains the $\beta$ dependence.
The mass-spectrum does not depend on $\beta$ and the theory becomes
a free one, as $\beta \rightarrow 0$. Therefore $S_{ab}(\te)Z_{ab}(\te)$
 goes to unity as $\beta\rightarrow 0$. If we build $Z_{ab}(\te)$ out of
$f_{\alpha}(\te)$ with $\alpha<0$, we introduce no new poles in the
physical sheet and are able to cancel all poles of $S_{ab}(\te)$ as
$\beta\rightarrow 0$. For the $Z(N)$ Toda models the $Z_{11}(\te)$-factor
for example is
\beq
    Z_{11}(\te)=f_{\beta}(\te)f_{-\frac{2}{N}+\beta}(\te).
\eeq
 Factorized $S$-matrices
describing Toda field theories with  structure corresponding to all
the simple Lie groups have been found recently \cite{TKEM,TODA}.

\subsection{A non-unitary series $A\pri(2N)$}

Finally we may generalize the non-unitary Lee-Yang edge singularity
to a whole series of non-unitary models\cite{CHIC}.This is achieved,
replacing the $Z(N)$-amplitudes $f_{\alpha}(\te)$ by
 $F_{\alpha}(\te)$ in equ.(\ref{eq:z2}),
such that they reduce to the Lee-Yang case for $N=1$.
In this way, we certainly obtain minimal amplitudes for a
set of $N$ self-conjugate particles:
\beq
   \esma{ab}{}=F_{\frac{|a-b|}{2N-1}}(\te)\left[
                       \prod_{k=1}^{min(a, b)-1} F_{\frac{|a-b|+2k}{2N-1}}(\te)
                               \right]^2
               F_{\frac{(a+b)}{2N-1}}(\te),
\label{eq:A2}
\eeq
with $1\leq a,b\leq N$. The mass-spectrum is given by
\beq
    m_a=\frac{ \sin\left( \frac{\pi a}{2N-1} \right) }
             { \sin\left( \frac{\pi}   {2N-1} \right) },\hspace{3em}
       a=1,2,\ldots,N-1.
\eeq
As opposed to the $Z(N)$-models, here the absence $Z(N)$-invariance
places no restrictions on simple poles to interpreted as particles.
The whole series therefore violates
unitarity \footnote{See however reference \cite{SMIRNOV}.}.
As we will see, these $S$-matrices belong to a series of non-unitary
conformal theories with negative central charge,
 perturbed by a relevant operator.

The investigation of possible conserved charges $P_s$ runs parallel
to our $Z(N)$ discussion, so that we don't repeat it here. One
finds\cite{CHIC}, not surprisingly for us at this stage,
 that an infinite number of $P_s$ exists for
\beq
      s =\;\;odd\;\;\neq 0\left( mod(2N+1) \right).
\eeq

Here also exist non-minimal solutions and their fundamental $Z$-factor
is given by\cite{TKEM}
\beq
      Z_{11}(\te)=F_{-\beta}(\te)F_{-\frac{2}{2N-1}}(\te).
\eeq
These non-mimimal models are now unitary, due to the $Z$-factors, which
change the sign of the relevant residues.

\sect{ The Thermodynamic Bethe Ansatz }{s3}

The Bethe Ansatz (BA) has been a powerful tool in analysing 2-dimensional
field theory and statistical mechanical models \cite{BA1}. It has been used to
diagonalize the hamiltonian of these models, to study their spectrum,
finite-size corrections etc.
Our purpose is to use the BA as a means to interpolate from the massive
$S$-matrix theory to the ultra-violet, short-distance fixed point,
in order to make contact with data from conformally invariant field
theory.

One way to use the BA is to start with the unrenormalized Hamiltonian
and to make an Ansatz for the wave functions
 of 'pseudo-particles' living on top of a
pseudo-vacuum. Their spectrum is obtained by imposing e.g.  periodic
boundary conditions on the wave functions. This yields the
so-called BA equations.  Bound states correspond to complex solution
in the rapidity of the BA equations.

These pseudo-particles have then to be used to  fill up
 a Dirac sea in order to construct the physical vacuum. The excitation
spectrum above this sea will correspond to the real particles of the theory,
whose scattering is described by the ( physical ) $ S $-matrix.

In the Thermodynamic Bethe Ansatz ( TBA ) \cite{TBA,BAZ} one
 deals with the real particles
from the very beginning, the rapidities therefore being real numbers in
the physical region. Bound states are treated on equal footing, which
is also in accord with the bootstrap principle.

Suppose now
we know the $S$-matrices of all the particles in the theory.
   We may then place them in a box, let them scatter till they are
in equilibrium at a particular temperature $T$. As we shall see, it is
 reasonable to assume,
that these equilibrium states will be described by BA wave functions,
since there is no production in our models.
As shown Al. B.  Zamolodchikov \cite{BAZ},
we may now extract finite size effects
and go in particular to the zero mass limit. In this way we obtain e.g.
the conformal anomaly ( or central charge ) $c$ of the ultraviolet
conformally invariant
limit of our massive field theory, establishing a beautiful connection
between factorizable $S$-matrices and conformal invariant field theories.

The importance of the central charge $c$\ , stems from the fact,  that its
knowledge is almost sufficient to pin down the conformal field theory
we are talking about. As will be explained in section \ref{s4} , it
can be extracted from finite-size effects arising in
a  massive Euclidean field theory defined on
a rectangle of size $L*R$ with , say ,  periodic boundary conditions
in the both directions - i.e. on a torus. In the following, we will
emphasize periodicity in the $R$-( {\em space}-)direction and talk about a
'vertical' cylinder of radius $R$.
If we slice the cylinder parallel to its base,  we may define the
transfer matrix $\hat T$ as the partition function of the small cylinder
$\Delta L*R$,  where we fix the configurations of the fields on
the neighboring bases \footnote{
                                We always assume some UV cutoff,  so that
                                we can count configurations and take the
                                limit at the end. We also choose
                                $\Delta L=1$. }.
In order to write the partition function  $Z(L, R)$  in terms of
the transfer matrix $\hat T$ along the $L-$({\em time}- )direction,
we introduce a complete set of states
$|n(L)>$ for each slice at position $L$. Due to periodic boundary conditions
in the vertical direction,  the states on the bottom and top
bases are the same. Labelling the slices by $1, 2, \ldots , 1$ , we get
\bdm
          Z(L, R)=\left(\prod_{j=1, 2, \ldots, 1} \sum_{n_j} \right)
                 <n_1|\hat T|n_2><n_2|\hat T|n_3>\ldots
                     <n_{L-1}|\hat T|n_1>=
\edm
\beq
                     \sum_{n_1}  <n_1|\hat T^L|n_1>=
                         {\cal T}r\; \hat T^L\;.
\eeq
The quantum hamiltonian of this theory is defined as
$\hat T=exp(-\hat H)$. Since the hamiltonian $\hat H$ is related to the
{\em time-time} component of the energy-momentum tensor, integrated
over space, we have
\beq
       \hat H=\frac{1}{2\pi}\int \hat T_{LL} dR,
\label{eq:TLL}
\eeq
where a factor $1/2\pi$ has been extracted for later convenience.

If $E_0(R) $ is the lowest eigenvalue of $\hat H$
( the ground state energy of our field theory ),
then for $ L \rightarrow \infty$ it will dominate the partition function :
\beq
              Z(L, R) \simeq exp( -LE_0(R)) .
\label{eq:ZE0}
\eeq

The central charge $c$ is now obtained
from  the free energy per unit length - which equals the
ground state energy $E_0$ of the two-dimensional conformal field theory -
\bdm
        F(R) =-\lim_{L\rightarrow \infty} \frac{1}{L} \ln Z(L, R)=E_0(R).
\label{eq:FE0}
\edm
 The TBA will extract $\tilde{c}$ and other
interesting quantities,  putting the system in a box
at finite temperature $T$.

Take as  'box' our cylinder. As $L\rightarrow \infty$,  we may look
at our system as a Euclidean quantum field theory ( EQFT )
 living on an infinite one-dimensional space with periodic
'time' in the $R$-direction \footnote{
        The period is $R$, the factor $2 \pi $ being
        absorbed in a rescaling of the energy.} .
 But this is exactly, what defines a
finite temperature Gibbs state (remember $k=1$!). Thus
$ Z(L, R)$ can be regarded as a EQFT at finite temperature $T=1/R$ with
free energy per unit length
\beq
     f(T)= -kT\, \frac{1}{L}\, ln\, Z(L, R)=-\frac{lnZ(L, R)}{LR}.
\label{eq:f}
\eeq
Therefore, comparing equs. (\ref{eq:ZE0}) and (\ref{eq:f}),
we finally get the relation
\beq
              E_0(R)=Rf(R).
\eeq

This establishes the desired link
between a quantity pertaining to a EQFT and a thermodynamic one accessible
to the TBA using $S$-matrix elements as input.

As we will show in  section \ref{s4} equ.(\ref{eq:Fc}),
the expansion of the free energy ( of the UV limiting theory )
in powers of $1/R$ is:
\beq
      F(R)=f_0\, R\, -\, \frac{\pi \tilde{c}}{6R}+O(1/R).
\label{eq:defc}
\eeq
Here $f_0$ is the non-universal bulk free energy per unit area of the
infinite plane,  which is chosen to vanish. Due to periodic boundary
conditions no term proportional to $R$ appears. $\tilde{ c}$ is related to
$c$ by $\tilde{c}=c-12\Delta_0$, where $\Delta_0$ is the lowest critical
dimension of the conformal field theory. For unitary theories
$\Delta_0=0$ and $\tilde{c} = c$.
In order to reach $c$ , we still have to take an UV limit
to get from our massive $S$-matrix theory to the conformally
invariant fixed point.
 From equ.(\ref{eq:ZE0}) we see that $E_0(R)$ has the dimension length$^{-1}$
and scaling arguments then imply
\beq
    E_0(R)=Rf(R)=\frac{1}{R}g(r)=-\frac{\pi\tilde{c}(r)}{6R},
\label{eq:scalf}
\eeq
where $r=R/\xi$. The correlation length $\xi$ is related to
 the mass of the lightest
particle ( in the thermodynamic limit, where Lorentz invariance holds and
where the momentum $P=2 \pi n/R$, $n=integer$ becomes a continuous
variable) as $\xi=m_{0}^{-1}$, so that $r=R m_0$.
In the UV limit, where  $r  << 1 $, $\tilde{c}(r)$
becomes the central charge $c=\tilde{c}(0)$.

Let us now set up the TBA equations, considering a EQFT at temperature
$T=1/R$ on a periodic one-dimensional space of length $L$.
For simplicity, let us start considering a system of $N$ identical particles
of mass m at equilibrium at temperature $T=1/R$.
Suppose we take particle a on a round-trip along the circle of length $L$.
On its way it will scatter against all other particles it encounters.
Let us look at this process in some detail \cite{ZZ}.
 If our system contains $N$
particles at positions $x_j$, there are $N!$ regions in
configuration space, where all particles are well separated -
$|x_i-x_j|\, >>\xi$ - and their mutual interaction can be
neglected~\footnote{Typically it will decay as $exp(-x/\xi)$. }.
In each of these regions, we can describe the system by an $N$-particle
wave function. Due to a scattering process - made up of two-particle
scatterings - , the system will pass from one region to another.
The wave function of the final region is now obtained by
multiplying the wave function of the initial region by the relevant
$S$-matrix ( $exp(-\imath\delta(\te) $ ).
On it's trip 'around the world' all the phase shifts will add up and due
to periodic boundary conditions the final wave function will be
identical to the one we started with. This has to be truefor anyone
of the $N$ particles. Thus we get :
\bdm
       e^{ \sum_{j} \imath (p_i + \delta_{ij})(x_i+L)} = e^{ \imath p_ix_i }
\edm
or \footnote{
     If the $S$-matrix is not diagonal, the wave functions
     will have internal indices,  on which the $S$-matrix acts.
     In this case the following equations turn into matrix equations.
                                                         }
\beq
        e^{\imath Lm\sinh (\te_i ) } \prod_{j\neq i}^N
                  S(\te_i -\theta_j ) = 1\:, \:
                                  i=1, 2, \ldots, N,
\eeq
or taking the logarithm
\beq
         L m\sinh \te_i +
          \sum_{j\neq i }^N \delta(\te_i -\te_j) =
          2\pi J_i \:,  \: i=1, 2, \ldots, N,
\label{eq:ba1}
\eeq
where $J_i$ are integers to be determined by equilibrium conditions. The
above {\em Bethe Ansatz equations } ( BAE ) reduce to the well known
equations in the free case
$ \delta_{ij}(\te) = 0 $, where they too determine the possible values
of the momenta $ p_i = m_i \sinh(\theta_i) = 2 \pi J_i/L $. In our
non-trivial, interacting situation they are a set of $ n$ coupled
transcendental equation for $ \theta_{ij} $.
The energy and momentum are given by
\beq
 \hfill E=\sum_{j=1}^N m\, \cosh\, \te_j,  \hspace{3em}
        P=\sum_{j=1}^N m\, \sinh\, \te_j,  \hfill.
\eeq

Notice that factorization
of the $S$-matrix once more has given us an unexpected surprise~:
how to extract off-shell information from on-shell data \footnote{
             In another step to implement this program, one obtains
             the form-factors from the on-shell information
              \cite{FF}.}.

In the following we will assume our Bethe wave functions to obey an {\em
exclusion} principle, i.e.  they are of {\em fermionic} type \cite{BAZ,TKEM}.
All our $S$-matrices are compatible with this condition. This  seems to be
a necessary requirement for the construction of a physical vacuum
in the usual Bethe-Ansatz \cite{IZKO}, although it is not necessary
for the implementation of the TBA \cite{MM}.

The wave function of identical bosons or fermions has in his case
to be anti-symmetric under their interchange. This implies the following
constraint on the $S$-matrix.
{}From the unitarity condition equ.(\ref{eq:unit}), we have $S_{aa}(0)=\pm 1$.
If $S_{aa}(0)=-1$, the wave function is anti-symmetric under interchange.
Therefore, in order to enforce an exclusion principle,
 we require $S_{aa}(0)=-1$ for bosons.
For fermion the situation is reversed and we require
$S_{aa}(0)=+1$ for fermions.

\subsection{The TBA equations in the thermodynamic limit}

To obtain information about the spectrum  of our theory,  we have
to investigate the roots of the BAE, which is a complicated system
of transcendental equations. But this system simplifies in the
thermodynamic limit, where $L \rightarrow \infty $ together with
the number of particles, which also grows $\sim L$.
The difference between adjacent solutions of equ.(\ref{eq:ba1})
goes to zero as $\te_i-\te_j \sim 1/m_0 L$. It is then reasonable
to define a density $J(\te)$ giving the number of real roots
in the rapidity interval $\theta$ and $\te + \Delta \te $.

Suppose now  that $ \{\te_1, \te_2, \ldots, \te_n\} $ is a self-consistent
\footnote{By this we mean that the $\te_j$, besides obeying the
          BAE, also satisfy the equilibrium conditions at
          temperature $T$ to be set up below. }
solution of equ.(\ref{eq:ba1}). Consider then the function
\beq
     J(\te) = \frac{m L}{2 \pi}\sinh \te
                 + \sum_{j}^N \delta(\te-\te_j),
\label{eq:JS}
\eeq
which - as will  be checked later on - is a monotonically increasing function
of $\te$.
Whenever $J(\te)$ passes through one of the integers $J_i$, the
corresponding $\te$ equals $\te_i$.
However there may
be integers $ J(\te) $ for which the corresponding $\te$ is {\em not}
in the set ${\te_1, \te_2, \ldots, \te_n} $. Such a $\te$ will be called a
{\em hole}.
Thus a natural definition of the density of {\em roots} and {\em holes}
is
\beq
        \rho(\te)=\frac{1}{L}\frac{dJ(\te)}{d\te}.
\label{eq:defrho}
\eeq

A linear integral equation for $J(\te)$ can now be obtained in the
limit $L\rightarrow \infty$, by first differentiating equ.(\ref{eq:JS})
\beq
     \frac{1}{L}\frac{dJ(\te)}{d\te}=\rho(\te)=
     \frac{m}{2\pi}\cosh \te + \sum_{i}^{N} \varphi(\te-\te_j),
\label{eq:JS1}
\eeq
where $\varphi(\te)$ is defined as
\beq
         \varphi(\te)\equiv \frac{d}{d\te}\delta(\te)
         = -\imath \frac{d}{d\te}\ln S(\te).
\eeq
We now make the replacement ( valid for sufficiently large $L$ )
\beq
  \sum_{j=1}^n f(\te_j)=\int d\te \rho(\te)f(\te) -
                         \sum_{j=1}^l f(\te_j)
\eeq
\bdm
                  = \int d\te \rho_r(\te)f(\te),
\edm
where $\te_1, \ldots, \te_l$ are the hole rapidities and
we introduced the density of {\em roots}  $\rho_r(\te)$.
Clearly the {\em hole} density is: $\rho_h=\rho - \rho_r$.

This transforms equ.(\ref{eq:JS1}) into
\beq
    \rho(\te)=\frac{m}{2\pi}\cosh\te+\varphi*\rho_r(\te),
\eeq
where we have introduced the convolution
\beq
     ( f \, *\,  \rho )(\te)=
                 \inti \frac{d\te\pri}{2 \pi} f(\te-\te\pri)
                                    \rho(\te\pri) .
\eeq

In general we may have more than one type of particles. Therefore introduce
densities for each type, labeled by  $a=1,\ldots,n$ : $\rho^{(a)}(\te)$, etc.
The integral equation for $ \rho^{(a)} = \rho_r^{(a)} + \rho_h^{(a)}$
becomes then :
\beq
       \rho^{(a)}(\te) =
        \frac{m_a}{2\pi} \cosh \te + \sum_{b=1}^n  \varphi_{ab} \, *\,
                                          \rho_r^{(b)}(\te),
             \hspace{1em}a=1, \ldots, n.
\label{eq:pbc1}
\eeq

As advertised, the very hard problem of solving $N$ coupled transcendental
equations has been converted to the solution of a simple linear integral
equation for the density $\rho_r^{(a)} $. However
we still need  a relation between $\rho_r^{(a)}$ and $\rho_h^{(a)}$.

In order to obtain the necessary additional information,  we impose the
condition of thermodynamic equilibrium. Thus we will calculate the
free energy per unit length $ f=e - Ts $, whose minimization will
provide what we want.

It is easy to compute the entropy of the distribution of particles
and holes on the circle $L$.
We have a large number of roots $ L\rho_r^{(a)}\, \Delta\te $ and holes
$ \rho_h^{(a)}\, \Delta\te $ in the interval $\Delta\te$. The number
of ways to distribute them in this interval
- according to our exclusion principle - is
\beq
  \frac{
       \left[L(\rho_r^{(a)} +\rho_h^{(a)})(\te) \,  \Delta\te
                                       \right]!              }
     {  [L\rho_r^{(a)}(\te) \,  \Delta\te]!
        [L\rho_h^{(a)}(\te) \,  \Delta\te]!                }.
\eeq
Taking the logarithm, we obtain the entropy per unit length:
\bdm
      s[\rho, \rho_r] = \sum_{a=1}^n s_a[\rho, \rho_r]
\edm
\beq
              =  \sum_{a=1}^n \inti \, d\te\,
       [(\rho_r^{(a)}+\rho_h^{(a)} )\ln( \rho_r^{(a)}+\rho_h^{(a)} ) -
       \rho_r^{(a)}\ln \rho_r^{(a)}- \rho_h^{(a)} \ln \rho_h^{(a)} ].
\eeq
       The energy per unit length is obviously given by
\beq
       e[\rho, \rho_r]=\sum_{a=1}^n \inti \, d\te \,
            \rho_r^{(a)}(\te) m_a \cosh \te.
\eeq
Let us introduce the function $\epsilon_{a}(\te) $ by
\beq
   \frac{\rho^{(a)}(\te) }{ \rho_r^{(a)}(\te) } =
            1 + e^{\epsilon_a(\te)}
\label{eq:defeps}
\eeq
and also
\beq
      L_a(\te) =  \ln ( 1 +  e^{-\epsilon_a(\te)})      .
\label{eq:el}
\eeq
Using this the entropy may easily be expressed as
\beq
    s[\rho_r, \epsilon_a]=
    \sum_{a=1}^n\inti \, d\te \rho_r^{(a)}(\te)
                [ \epsilon_a(\te) + (1 + e^{\epsilon_a(\te) } )
                           \ln( 1 + e^{-\epsilon_a(\te) } )].
 %
\label{eq:entr}
\eeq
The equilibrium distribution at temperature $T$ is now given by
minimizing the free energy $f=e-Ts$ with respect to $\rho_r^{(a)} $,
subject to the periodic boundary condition equ.(\ref{eq:pbc1}) relating
$\rho^{(a)} $ to $\rho_r^{(a)} $.

Thus using
\bea
  \frac {   \delta s }{ \delta \rho_r } &=&
      \frac{ \delta s }{\delta \rho_r} +
      \frac{ \delta s}{ \delta \rho } \frac{ \delta \rho}{\delta \rho_r }
       \nonumber \\
    &=& \ln \frac{ \rho-\rho_r }{\rho_r }
        + \frac{\delta\rho}{\delta\rho_r} \ln \frac{\rho}{\rho-\rho_r}
\eea
and
\beq
     \frac{\delta\rho(\te)}{\delta\rho_r(\te\pri)}
     =\varphi(\te-\te\pri),
\eeq
coming from the periodic boundary condition, we get finally get the
following extremum condition:
\beq
            \frac{m_a}{T} \cosh \te =
      \epsilon_a(\te) + \sum_{b=1}^n \varphi_{ab} \, *\,  L_b (\te).
\label{eq:extr}
\eeq
Note that this condition depends only on $\epsilon_a(\te) $.
The solution of this equation for $\epsilon_a(\te)$ will always
be real. From equ.(\ref{eq:defeps}) we see that this implies
$\rho^{(a)}(\te)>0$, since $\rho^{(a)}_r(\te)$ is positive by
definition. This in turn results - see equ.(\ref{eq:defrho}) - in
$J^{(a)}(\te)$ being a monotonically increasing function of $\te$,
as announced.

The equilibrium free energy can now be obtained by eliminating $\rho_h$,
using the periodic boundary condition equ.(\ref{eq:pbc1}) and
\makebox{  $\rho_h/\rho_r=e^{\epsilon(\te)} $ } in the expression for
 $s[\rho_r, \rho_h] $. Using the extremum condition
equ.(\ref{eq:extr}) then yields the free energy :
\beq
   f(\te) = - T\sum_{a=1}^n \inti \frac{d\te}{2 \pi}
                m_a\cosh\te L_a(\te).
\label{eq:freee}
\eeq
Usually numerical work will now be envolved in order to compute this
equilibrium free energy. But a number like the conformal anomaly can
be extracted analytically.

To accomplish this we have to go to the UV limit, where the masses
 may be neglected :
$ r=m/T \rightarrow 0 $.
Let us first take the limit of the extremum condition equ.(\ref{eq:extr})
\beq
            \hat{m}_a r \cosh \te =
      \epsilon_a(\te) + \sum_{b=1}^n \varphi_{ab} \, *\,  L_b (\te),
\label{eq:extr1}
\eeq
where we have introduced dimensionless mass ratios
$\hat{m}_a=m_a/m_0=m_a\xi$. As $r\rightarrow 0$, $\te$ has to go to $\infty$
for the l.h.s. to give a finite contribution, in which case it behaves as
\bdm
      \hat{m}_a r \cosh\te \sim \hat{m}_a \frac{r}{2} e^{\te}=
      \hat{m}_a \exp (\te - \ln \frac{2}{r} ).
\edm
We therefore make the shift $\te \rightarrow \te+\ln(2/r)$ to get the
following equation:
\beq
     e^{\te}=\tilde{\epsilon}_a(\te)+
        \sum_{b=1}^n ( \varphi_{ab} \, *\,  \tilde{L}_b )(\te),
\label{eq:assextr}
\eeq
whose solution provides the $r${\em -independent} functions
\beq
          \tilde{\epsilon}_a(\te)\equiv \epsilon( \te+\ln (2/r) )
\eeq
\bdm
          \tilde{L}_b(\te) \equiv L_b( \te + \ln (2/r) ).
\edm
Solving numerically the extremum condition for $\epsilon_a(\te)$,
one realizes that this function equals the constant
 $\epsilon_a$ in the range $ -\ln(2/r) << \te << +\ln(2/r) $  and eventually
grows exponentially, so that
$\tilde{\epsilon}_a(\te) = \epsilon_a$ , except at $\infty$,
for $r\rightarrow 0$ \cite{BAZ}.
Similarly
$\tilde{L}_b(\te)$ interpolates between $\epsilon_a$ at $\te=0$ and
- through a double exponential decay - $0$ for $\te \neq 0$.

If we now use equ.(\ref{eq:scalf}) and take the limit $r \rightarrow 0$
in equ.(\ref{eq:extr1}), we get a compact formula for $\tilde{c}(0)$:
\beq
   \tilde{c}(0)=\frac{3}{\pi^2}\sum_{a=1}^n \hat{m}_a
                    \inti d\te \tilde{L}_a(\te)e^{\te},
\label{eq:BAc}
\eeq
which is very convenient for a numerical solution.

A perhaps more explicit formula for $\tilde{c}(0)$ may also be obtained
in terms of Roger's dilogarthmic function \cite{BARE}.
It is  more convenient to use
the entropy equ.(\ref{eq:entr}) and to get the $\tilde{c}(0)$ from
\beq
        S= \frac{ \pi \tilde c(r) }{3}T \,  L +  O(T^2),
\label{eq:centr}
\eeq
which is equivalent to equ.(\ref{eq:defc}) \footnote{
              We have $f(T)=e-Ts=\ldots-T^2(\pi\tilde{c}/6)$. But
              $s=-\partial f/\partial T=T(\pi\tilde{c}/6)+O(T^2)$ and
              equ.(\ref{eq:centr}) follows.}.

Again the entropy $s[\rho_r, \epsilon_a] $ vanishes in the
limit $r \rightarrow 0 $, as long as the limits of the integral over
$ \te $ are finite. Now check the behavior of the integrand. On
one hand taking the
 derivative  with respect to $ \te $ of the extremum condition
equ.(\ref{eq:extr}) we get
\beq
     \frac{d\epsilon_a(\te)}{d\te}\simeq
          \frac{m_a}{T}\sinh \te \ \simeq
          \frac{m_a}{2T} e^{|\te|}sgn\te,  \:\:|\te| \rightarrow\infty.
\eeq
On the other hand,
since $ \varphi(\te)=0 + O(e^{-\te})$ as
$ \te \rightarrow \infty $,  we get for the asymptotic behavior of
$ \rho_r^{(a)} $ from the BAE equ.(\ref{eq:pbc1}) :
\beq
  \rho^{(a)} \simeq \frac{m_a}{4 \pi}e^{|\te|)}
\eeq
and consequently
\bea
   \rho_r^{(a)} & \simeq & \frac{m_a}{4 \pi}\frac{e^{|\te|}}
                                        {1+e^{\epsilon_a(\te)}}
                         \nonumber \\
                & \simeq & \frac{T}{2 \pi} \frac{d\epsilon_a(\te)}
                                                {d\te}
                           (1+e^{\epsilon_a(\te)})^{-1}(sgn\te).
\eea
Substituting this into equ.(\ref{eq:entr}) for the entropy, we get
\beq
  \lim_{r\rightarrow 0} s[\rho_r]  =
   \frac{T}{2\pi} \sum_{a=1}^n\inti\, d\te \frac{d\epsilon_a(\te)}{d\te}
     (sgn\te)[\ln(1+e^{\epsilon_a(\te)})-
              \frac{\epsilon_a(\te)}{1+e^{-\epsilon_a(\te)})} ].
\eeq
Here we change variables $\te\rightarrow \epsilon_a$ and use
$\varphi_{ab}(\te)=\varphi_{ab}(-\te) $, which follows from the unitarity
of the $S$-matrix ,  to show that $\epsilon_a(\te)$  is even in $\te$.
Finally changing variables again from $\epsilon_a$ to
$f(\epsilon_a)=1/(1+exp(\epsilon_a))$,  we get :
\beq
  \lim_{r\rightarrow 0} s[\rho_r]  =
   \frac{T}{2\pi} \sum_{a=1}^n\int_{f[\epsilon_a(0)]}^{f[\epsilon_a(\infty)]}
             \, dy\, [ \frac{\ln y}{1-y} + \frac{\ln (1-y)}{y} ] .
\eeq
The upper limit of the integral equals $0$, since
$\epsilon(\te)\rightarrow\infty$ there. The lower limit may be
obtained from the extremum condition equ.(\ref{eq:extr}) , setting there
$ \te\ \rightarrow 0 $ and $ r \rightarrow 0 $:
\beq
   0=\epsilon_a(0)+\sum_{b=1}^{n}\inti\frac{d\te\pri}{2\pi}
                    \varphi_{ab}(\te-\te\pri)L_{b}(\te\pri).
\eeq
However
$\varphi_{ab}(\te)$ decreases
exponentially off the origin and we get an equation for
$\epsilon_a\equiv\epsilon_a(0)$
\beq
          e^{\epsilon_a}= \prod_{b=1}^n ( 1+e^{-\epsilon_b})^{N_{ab}}
        \hspace{2em}a=1, \ldots, n,
\label{eq:Neps}
\eeq
where the symmetric matrix $N_{ab}$ is:
\beq
   N_{ab} \equiv -\inti \frac{d\te}{2\pi}\varphi_{ab}(\te)=
          -\frac{1}{2\pi}[\delta_{ab}(+\infty)-\delta_{ab}(-\infty)].
\label{eq:Nab}
\eeq
  Introducing Rogers' dilogarithmic function $L(x)$ as
\beq
       L(x)=-\frac{1}{2} \int_0^x\,  dy\,
                \left[ \frac{\ln y}{1-y}  + \frac{\ln (1-y)}{y} \right],
\eeq
we may finally express \~c, using the finite-size scaling formula
equ.(\ref{eq:centr}),  as
\bdm
   \tilde c=\sum_{a=1}^n \tilde c_{a}(\epsilon_a),
   \hspace {5em}where
\edm
\beq
    \tilde c(\epsilon) =
 \frac{6}{\pi^2}\, L\left(\frac{1}{1+e^\epsilon}\right)=
              \frac{6}{\pi^2}\, L\left( e^{-\epsilon}\right)
\label{eq:cc}
\eeq
Each particle species contributes with $\tilde{c}_a(\epsilon_a)$ to the
total central charge. The function $\tilde{c}_a(\epsilon_a)$ is strictly
monotonically decreasing, approaching $0$ as $\epsilon\rightarrow\infty$
and $\tilde{c}_a(0)=1$.
In order to obtain $\tilde{c}$ the non-linear equation (\ref{eq:Neps})
has to be solved.

\subsection{Computation of the central charge}

Let us try things out in some simple cases, although there exist
solutions for whole sets of models\cite{TKEM}.

 {\bf 1) Ising model ( $h=0,T\neq T_c$ )}

The Ising model ( or $Z(2)$ model ) is a free fermion theory with
$\varphi_{ab}=const$ and
$N_{ab}=0$. We may start directly from equ.(\ref{eq:freee}) for the
free energy. Thus we have to compute
\bdm
  E_0(r)=Rf(R)=-\frac{\pi}{6R}c=
\edm
\bdm
        -\frac{1}{R}\inti \frac{d\te}{2\pi} r\, \cosh\te
                  \ln(1+e^{-\epsilon(\te)})
\edm
in the limit $r\rightarrow0$. With $\epsilon(\te)=r\cosh\te\simeq
re^{|\te|}/2$, we get
\bdm
               r\inti \frac{d\te}{2\pi} \cosh\te
                  \ln(1+e^{-\epsilon(\te)})
              \simeq 2 \int_0^{\infty} \frac{d\te}{2\pi}
                \epsilon(\te)\pri \ln(1+e^{-\epsilon(\te)})   \simeq
           \int_0^{\infty} d\epsilon\,  \ln(1+e^{-\epsilon})
            =\frac{\pi}{12}.
\edm
     With this result we obtain,  as expected $c=1/2$. As a matter of fact,
we may obtain an expansion of $R\, E_0(R)$ in powers $r$ and the result
compares well with numerical finite-size calculations \cite{SAZA}.

{\bf 2) Lee-Yang edge singularity\cite{BAZ}}

In section \ref{s1}, we found the $S$-matrix
\beq
      \esma{11}{}=F_{2/3}(\te).
\eeq
{}From the properties equ.(\ref{eq:F}), we have
\bdm
  F_{\alpha}(\te)=f_{\alpha}(\te)f_{\alpha}(\imath\pi-\te)=
   f_{\alpha}(\te)(-1)f_{1}(\te)f_{1-\alpha}(\te).
\edm

Since all our $S$-matrices are products of the $f_{\alpha}(\te)$'s,
their contributions to the phase shifts and the matrix $N_{ab}$ of
equ.(\ref{eq:Nab}) has the structure $N_{ab}=\sum_j\, N[f_{\alpha_j}]$.
For $N[f_{\alpha}]$ we get $N[f_{\alpha}]=(1-|\alpha|)sgn(\alpha)$ for
 $\-1<\alpha\leq1$ and $ sgn(0)\equiv0$.

Consequently we get $N[F_{2/3}]=1$ and equ.(\ref{eq:Neps}) becomes
\beq
      e^{\epsilon_1}=1+e^{-\epsilon_1},
\eeq
whose solution is
\beq
     \epsilon_1=\log[(\sqrt{5}+1)/2]
\eeq
and $\tilde{c}$ can be extracted form equ.(\ref{eq:cc}) to give
$   \tilde{c}= 2/5 $. If we take the value of $\Delta_0$ from
conformal field theory: $\Delta_0=-2/5$,
we get for the central charge $c=-22/5$. Notice the negative value
for $c$, as appropriate for non-unitary models.

{\bf 3) Z(N)-models}

Now let us look at $Z(N)$- (also called $A_{N-1}$) models
 in some detail.
The complete two-particle $S$-matrix was written in equ.(\ref{eq:z2}):
\beq
   \esma{ab}{}=\{ |a-b|/N\}\left[
                       \prod_{k=1}^{min(a, b)-1} \{\frac{|a-b|+2k}{N}\}
                               \right]^2
               \{(a+b)/N\},
\eeq
where $a, b=1, 2, \ldots, N-1$ and we used the shorthand $\{\alpha\}\equiv
f_{\alpha}(\te)$.

Let us check  $Z(3)$. We have : $N_{11}=N_{22}=N[f_{2/3}]=1/3$ and
$N_{12}=N_{21}=N[f_{1/3}]+N[f_{3/3}]=2/3$. Therefore we have to
solve the equs.(\ref{eq:Neps}):
\beq
   e^{\epsilon_1}\equiv x_1=(1+e^{-\epsilon_1})^{1/3}
                          (1+e^{-\epsilon_2})^{2/3}
\eeq
\bdm
   e^{\epsilon_2}\equiv x_2=(1+e^{-\epsilon_1})^{2/3}
                          (1+e^{-\epsilon_2})^{1/3}.
\edm
Their solution is $ x_1=x_2=2\cos(\pi/5)$.
If we now use the sum-rule \cite{KIRE}
\beq
    \sum_{k=2}^{n+1}L\left( \frac{\sin^2(\pi/n+3)}
                                 {\sin^2(\pi k/n+3)} \right )
             =L(1) \frac{2n}{n+3}
             =\frac{\pi^2}{6} \frac{2n}{n+3}\;,
\eeq
we obtain $c$ from equ.(\ref{eq:cc})  as
\beq
       c=\frac{6}{\pi^2}\sum_{a=1}^{2}  L(e^{-\epsilon_a})
               = \frac{6}{\pi^2}\frac{\pi^2}{6}\frac{4}{5}
               =\frac{4}{5},
\eeq
which is the expected value $c\left(Z(3)\right)=4/5$.

For general $N$ the solution of equs. (\ref{eq:Neps}) is \cite{BARE}
\beq
    e^{\epsilon_a}= \frac{\sin \frac{a\pi}{N+2} \sin \frac{(a+2)\pi}{N+2}}
                         {\sin^2 \frac{\pi}{N+2} }.
\eeq
and the same sum rule plugged into equ.(\ref{eq:cc}) yields :
\beq
     c\left(Z(N)\right)=\frac{2(N-1)}{N+2},
\eeq
which is the central charge of the $Z(N)$ parafermion models.

   The non-minimal models describing Toda field theories have, as
shown in section~\ref{s2}, a free UV limit and therefore their
value of $c$ is simply the number of bosonic fields, as can
be explicitly verified~\cite{TKEM}.

For further applications of this scheme see e.g. ref.\cite{Mox}.

{\bf 3) The non-unitary series}

Following the reasoning exhibited for the Lee-Yang edge singularity
and the $Z(N)$ models, we obtain:
\beq
    e^{\epsilon_a}\left(A\pri(2N)\right)=
    e^{\epsilon_a}\left(Z(2N)\right).
\eeq
{}From the same sum-rule used in the $Z(N)$ case we get:
\beq
    \tilde{c}\left(A\pri(2N)\right)=
    \frac{1}{2}\tilde{c}\left(Z(N)\right)=\frac{2N}{2N+3}.
\eeq
Borrowing $\Delta_0=-\frac{2(N-1)(6N-1)}{2N+1}$, we get for the central
charge:
\beq
      c\left( A\pri(2N)\right)=-\frac{2(N-1)(6N-1)}{2N+1}.
\eeq

At this point we have completed the first of our connections between
a massive $S$-matrix theory and it's UV-limiting fixed point data.It gives
us confidence, that the plausibility assumptions made in the previous section,
are indeed correct.In order to further clarify this point, we will dig into
this UV theory in more detail in the next section. \sect{ Lightning Overview of
   Conformal Invariance }{s4}
   The purpose of this section is to provide a minimum of
vocabulary for the newcomer or a reminder for the rusty
practitioner, but it is not a substitute for the excellent
review articles and books on the subject \cite{CR,ZAMrev}.

\subsection{ Conformal transformations }

Conformal transformations are coordinate transformations
$ x_{\mu} \rightarrow y_{\mu}(x) $, which preserve
the angle between two arbitrary vectors at any point, although
 their length may change:
\beq
  dy_{\mu}dy^{\mu}=
        \frac{\partial y_{\rho}}{\partial x^{\mu} }
        \frac{\partial y^{\rho}}{\partial x^{\nu} }
        dx^{\mu}\,dx^{\nu}
        \equiv  \rho (x)\,dx_{\mu}\,dx^{\mu}.
\label{eq:conf}
\eeq
 A special case are dilatations, which scale all lengths up by
a factor $ \rho (x)=\lambda$.
Under mild assumptions - for  D=2 a discrete spectrum of dimensions -
a scale invariant quantum
field theory is also conformally invariant \cite{polch}.
Therefore in Statistical
Mechanics the long distance behavior at a critical point,
 which is scale invariant due to a diverging correlation length,
 is described by a conformally invariant
Euclidean field theory \cite{poly3}.

Although for dimensions $D>2$, the conformal group is
finite-dimensional,  for $D=2$ it involves an infinte number
of parameters, which entails powerful restrictions on the theory.

Consider then a Euclidean Field theory in the 2-dimensional space of
coordinates $x_1,x_2$ \footnote{
         The Minkowskian space-time would have coordinates
         $x^1$ and $x^2=it$.
                               }.
We,  very conveniently, introduce complex ( light-cone ) coordinates
\beq
          z \,=\,x^1+\imath \,x^2; \;
     \rev{z}\,=\, x^1-\imath \,x^2.
\eeq
Then any transformation
\beq
       z\rightarrow z\pri =f(z)\;;
       \rev{z} \rightarrow \rev z\pri=\rev{f}(\rev{z})\;,
\eeq
where $f(z)$ and $\rev{f}(\rev{z})$ are differentiable functions,
 are conformal trans\-for\-ma\-tions satisfying equ.(\ref{eq:conf}).
We will consider $z$ and $\rev{z}$ to be independent complex variables.
However, since physical correlation
functions live in the real space $R^2$, we eventually have to take
into account that $\rev{z}$ is the complex conjugate of $z$:
$\rev{z}=z^*$, when constructing {\em physical} correlation functions.

Under the finite conformal transformation \footnote{In doing
            this we want to stay within the conformal plane. Thus by
            {\em finite} we mean  {\em global } transformations of the form
             $ z\pri=\frac{az+b}{cz+d}$ with $ad-bc=1$, which are the only
             one-to-one mappings of the complex plane onto itself.}
$ z \rightarrow z\pri=f(z),\rev z\rightarrow \rev z\pri=\rev f(\rev z)$
the Jacobian is
\beq
         \frac{\partial(z\pri,\rev z\pri) }
               {\partial(z,z\pri) }
           = f\pri(z)\;\rev f\pri(\rev z) .
\eeq
Tensor fields will transform with some powers of $ f\pri$ and $\rev f\pri$.
We will define {\em conformal fields of weight}
$(h,\rev h) $ to transform as
\beq
      \phi (z,\rev z)\rightarrow \phi\pri(z,\rev z)=
         f\pri(z)^h\;\rev f\pri(\rev z)^{\,\rev h}
         \phi (f(z),\rev f(\rev z) ).
\eeq
The scaling dimension is $\Delta=h+\rev h$, whereas the spin is
$s=h-\rev h$. We also want to consider $f(z)$ and $\rev f(\rev z) $ to be
an  arbitrary, but
infinitesimal change of coordinates \footnote{
          Let us mention a delicate point here. Infinitesimal, everywhere
          analytic functions do not exist, since they are bound to
          develop singularities somewhere, if they are nontrivial.
          Thus we  consider a bounded domain, which excludes
          the singularities.}:
 $\delta z=\epsilon(z),\delta \rev{z}=\rev {\epsilon}(\rev z)$.
The above equation then becomes
\beq
     \delta \phi(z,\rev z)=( \epsilon(z)\partial_z+h\,\epsilon\pri(z)
         +  \rev \epsilon(\rev z)\partial_{\rev z}+
           \rev h\,\rev {\epsilon}\pri(z) )
                            \phi(z,\rev z).
\eeq
and fields transforming this way are called {\em primary}.
All fields we will deal with are primary, except the
energy-momentum tensor.

\subsection {The energy-momentum tensor}

The energy-momentum tensor $T^{\mu \nu} $ is the generator of
space-time symmetries and as such is symmetric and conserved:
$T^{\mu \nu}=T^{\nu \mu},\partial^{\mu}T^{\mu \nu}=0$.
Due to these properties it is a prominent object in  the study of conformal
transformations in field theory.
If we consider the infinitesimal coordinate transformation - not
necessarily conformal ! -
\beq
     x^{\mu} \rightarrow x^{\mu} + \epsilon^{\mu}(x) ,
\label{eq:eps}
\eeq
then we may define $T^{\mu \nu}$ as inducing the following change
in the action
\footnote{
          For example,  in the case of a free massless boson with
          Lagrangian density ${\cal L}=\frac{1}{2}(\partial_{\mu}\phi)^2$,
          we get with this definition
          $ T_{\mu \nu} =  -\partial_{\mu}\phi\partial_{\nu}\phi
                          +\frac{1}{2}\delta_{\mu \nu}
                           \partial_{\alpha}\phi\partial^{\alpha}\phi.$
          }:
\beq
   \delta S=-\frac{1}{2\pi}\;\int\, T_{\mu \nu}
          \partial^{\mu} \epsilon^{\nu}\; d^2 x.
\eeq
In terms of correlation functions this is equivalent to the
identity
\bdm
 \sum_{j=1}^n\,<\phi_1(x_1)\ldots\delta_{\epsilon}\phi_j(x_j)\ldots
         \phi_n(x_n)>=
\edm
\beq
 -\int \frac{d^2x}{2 \pi }\, \partial_{\mu}\epsilon_{\nu}(x)
   <\;T^{\mu \nu}(x) \phi_1(x_1)\ldots\phi_n(x_n)\;>,
\label{eq:delT}
\eeq
where $\delta_{\epsilon}\phi(x)$ is the variation of $\phi(x)$ under
the transformation equ.(\ref{eq:eps}) \footnote{
         Fast derivation ! Define correlation functions by the
         functional integral (or Gibbs average on a lattice in
         statistical mechanics ):
\bdm
       < \phi(x)\ldots>=\frac{ \int {\cal D}\phi \;\phi (x)\ldots
                                    e^{-S[\phi]} }
                             { \int {\cal D}\phi\;e^{-S[\phi]} } .
\edm
        In the numerator of the functional integral on the r.h.s. change the
        dummy variable $\phi(x) \rightarrow \phi(x) +
          \delta_{\epsilon}\phi (x) $. Expand
        everything to first order in  $\delta_{\epsilon} \phi(x)$,
        getting two terms:
        one from $\phi(x)$ and another from the variation of the action
        $S[\phi+\delta_{\epsilon}\phi]=S[\phi] + \int d^2x/(2\pi) T^{\mu \nu}
           \partial_{\mu}\epsilon_{\nu}(x) $. The zeroth order term
       cancels the l.h.s. and we get the above Ward identity.
                                                }.
This is a very useful definition
of $T_{\mu \nu}$, if we don't know the Lagrangian of the theory we are
studying.  If we are after exactly integrable models it is often more
profitable to shift the emphazis to other types of structures like
symmetry properties, operator
product expansions, fusion rules  etc.\~, the corresponding
Lagrangian being determined a posteriori.

If we introduce the fields $T=(T^{11}-T^{22}+2iT^{12})/2$ and
$\rev T=(T^{11}-T^{22}-2\imath T^{12})/2$, then the conservation equations
$T^{\mu \nu}=T^{\nu \mu},\partial^{\mu}T^{\mu \nu}=0$ become
\beq
      \partial_{\rev z}T=-\partial_{z}\Theta/2\;;
      \partial_{ z}{\rev T}=-\partial_{\rev z}\Theta/2;
\eeq
where $\Theta=T^{11}+T^{22}$ is the trace of the energy-momentum tensor.
In a conformally invariant theory this trace vanishes and $T$ and $\rev T$
are analytic functions of $z$ and $\rev z$ respectively:
\beq
          T=T(z),\; \rev T=\rev T (\rev z).
\label{eq:Tz}
\eeq
The 2-dimensional problem is thus reduced to effectively two
one-dimensional problems
and it is this feature, which makes conformally invariant models
exactly soluble. Equs.(\ref{eq:Tz}) mean that the correlation functions
\beq
     <T(z) \phi_1(z_1,{\rev z}_1)\ldots\phi_n(z_n,{\rev z}_n)>
\eeq
are single-valued analytic functions with singularities (at most poles
of finite order) only at the points $z_1,z_2,\ldots,z_n$  and similarly
for $\rev T(\rev z)$.

To see this
integrate the r.h.s. of equ.(\ref{eq:delT}) over a domain excluding
small regions around the points $x_j$ and choose for $\epsilon(z),
\rev \epsilon(\rev z)$ functions vanishing sufficiently fast at
infinity to allow partial integrations. In this domain
$\partial_{\mu}T^{\mu \nu}$ will be zero and only the surface terms
around the points $x_j$ survive. Putting the origin at one such point,
and choosing for the excluded region a small circle of radius
$\rho$, yields the surface term
 $  \int d\sigma_{\mu}T^{\mu \nu}\epsilon_{\nu}$. With
\bdm
  d\sigma_{\mu}=d^2x\frac{\partial F(x)}{\partial x^{\mu} }
                \delta ( F(x) -\rho)\, ,\,
          F(x)=x_1^2+x_2^2
\edm
we get
\bdm
           \int d\sigma_{\mu}=\oint_{\rho} d\theta x_{\mu}.
\edm
Transforming from
rectangular to complex coordinates, we obtain for the  r.h.s.
of equ.(\ref{eq:delT}) the result:
\beq
  \int d\sigma_{\mu}T^{\mu \nu}\epsilon_{\nu}=
   \frac{1}{2 \pi \imath }\left\{
    \oint dz\, T(z)\epsilon(z)+\oint d\rev{z}\, \rev{T}(\rev z)
                                \rev{\epsilon}(\rev z) \right\}.
\label{eq:QT}
\eeq
    Thus the Ward identity for $T(z)$ (and a similar one
for $\rev T(\rev z)$ ) becomes :
\bdm
      \sum_{j=1}^n\,<\phi_1(z_1,\rev z_1)\ldots
      \delta\phi_j(z_j,\rev z_j)\ldots>=
\edm
\bdm
      \sum_{j=1}^n\,\left( \epsilon(z_j)\partial_{z_j}+
         h_j \epsilon\pri(z_j) \right)
      <\phi_1(z_1,\rev z_1)\ldots\phi_n(z_n,\rev z_n)>=
\edm
\beq
    \oint_{\cal C} \frac{dz}{2 \pi \imath} \epsilon(z)
<T(z)\phi_1(z_1,\rev z_1)\ldots\phi_n(z_n,\rev z_n)>,
\label{eq:wardT}
\eeq
where ${\cal C}$ encircles the points $z_1,\ldots,z_n$
 once in the positive sense.
This statement is equivalent to - as can be seen multiplying
equ.(\ref{eq:ward2}) by $\epsilon(z)$ and integrating
along ${\cal C}$ -
\bdm
    <T(z)\phi_1(z_1,\rev z_1)\ldots\phi_n(z_n,\rev z_n)>=
\edm
\beq
    \sum_{j=1}^n \left(
      \frac{h_j}{(z-z_j)^2}+\frac{1}{z-z_j}\partial_{z_j}\right)
    <\phi_1(z_1,\rev z_1)\ldots\phi_n(z_n,\rev z_n)>
\label{eq:ward2}
\eeq
 and a similar equation for $\rev T(\rev z)$.
These relations express the transformation properties of primary
fields under conformal transformations.

Surprisingly the energy-momentum tensor itself is not a primary field.
As a matter of fact, in a non-trivial theory the two-point function
$<T(z)T(0)>$ cannot
      vanish  and since the dimension  of $T(z)$ is $h=2$
\footnote{This can already be seen from equ.(\ref{eq:ward2}).
          Or : the generator of translations by a vector
          $a_{\mu}$ is $exp(\imath a_{\mu}P^{\mu})$.
          Therefore $P^{\mu}=\int T^{\mu \nu}d\sigma _{\nu}$
          has dimension one and $T^{\mu \nu}$ has dimension two.},
scale invariance implies
\beq
     <T(z)T(0)>=\frac{c/2}{z^4}.
\eeq
The dimensionless constant $c$ is called {\em central charge}
or {\em conformal anomaly} and it's knowledge characterizes to a large
extent a conformal model.
$c$ has been normalized such that the free scalar massless boson
has $c=1$.
The analog of equ.(\ref{eq:ward2}) for $T(z)$ itself now becomes
\footnote{
      This expression is to be understood as occurring
      inside a correlation function. The ellipsis stands for
      terms, which are regular as $z_1\rightarrow z_2$,
      generated when $T(z_1)$ hits other fields of the
      correlation function. The reader, who is uneasy with
      this kind of operator statements should take
      refugee in Furlan et al.\cite{CR}.}
\beq
    T(z_1)T(z_2)=\frac{c/2}{(z_1-z_2)^4}+
     \frac{2}{(z_1-z_2)^2}T(z_2)+\frac{1}{(z_1-z_2)}\partial_1 T(z_2)
     +\ldots.
\label{eq:TT}
\eeq

Equ.(\ref{eq:TT}) translates into the following transformation
law for $T(z)$:
\beq
     \delta T(z) = \frac{1}{12}c\epsilon ^{'''}(z)
                   +2\epsilon\pri(z)+\epsilon(z)\partial _zT(z).
\eeq

It is not completely trivial to integrate this
quation to obtain the transformation law for
finite conformal transformations. We state only the result \cite{CR}:
\beq
         T(z)= (f\pri(z\pri))^2 T\pri(z\pri)+\frac{c}{12}S(f,z),
\eeq
where the {\em Schwartzian} derivative is
\bdm
    S(f,z)=\frac{\partial_zf\,\partial_z^3f\,-\frac{3}{2}(\partial_z^2f)^2}
                {(\partial_zf)^2}.
\edm

This is a very useful result, for it permits to connect informations
pertaining to different geometries. For arbitrary $f(z)$,  not of the form
$f(z)=(az+b)/(cz+d)$, the conformal plane will be mapped into a different
geometrical domain. We now
 broaden the concept of conformal
invariance to include invariance under
this {\em active} transformation. Let us,  for example,
conformally map the complex z-plane ( without the origin ) onto a periodic
{\em horizontal} strip of width $R$ in the $w=u+\imath v$-plane
 by $z=exp(2\pi wR)$
 and analogously for $\rev z$.
We now postulate that objects in the $z$-plane go, via this
mapping,  over to the corresponding objects in the strip. Applying this
to the energy-momentum tensor, we get
\beq
     T_{strip}(u)=(\frac{2 \pi}{R})^2\,[T_{plane}(z)z^2-c/24].
\eeq

In particular, if we now take the expectation value of $T(z)$,
 since $<T_{plane}(z)>$ has been renormalized to zero,  we find
\beq
  <T_{strip}(u)>=-\frac{c}{24}\frac{2 \pi}{R}^2.
\eeq
This formula allows us to measure the central charge $c$ exploring
finite size effects in the statistical mechanical version of our
Euclidean quantum field theory, where $c$ is obtained via the
free energy.

In fact the variation of the free energy is given by a
formula analogous to equ.(\ref{eq:delT}) with no fields on the l.h.s.~:
\beq
   \delta ln Z=-\int _{{\cal D}}\frac{d^2u}{2 \pi }
                < T_{\mu \nu}(u,\rev u)>\partial_{\mu}\epsilon_{\nu},
\label{eq:delZ}
\eeq
where the integration domain ${\cal D}$ is the infinite strip.
Choose a transversal dilatation by $\delta\eta$ of the strip~:
$\epsilon_1=0,\epsilon_2=u_1\delta\eta$.
We get $T_{\mu \nu}\partial_{\mu}\epsilon_{\nu}
=(T(u)+\rev T(\rev u))\delta\eta$. The accessible statistical
mechanical observable is the free energy per unit (vertical) length
with suitable boundary conditions in the vertical direction
\beq
      F(R)=-\lim_{M\rightarrow \infty}\frac{1}{M}\,lnZ(R,M).
\eeq
Under the transversal dilatation $F(L)$ changes by
$\delta F(L)=R\delta\eta\,dF(R)/dR$ so that from equ.(\ref{eq:delZ})
follows, that
\beq
         dF/dR=(2 \pi/R^2)(c/12).
\eeq
Using $F(\infty)=0$,  we get
\beq
     F(R)=-c\frac{\pi}{6}\frac{1}{R}.
\label{eq:Fc}
\eeq
This equation holds true, if all operators have positive dimensions, as is the
case of unitary models. As we shall see,
for non-unitary models, we have at least one operator
 with negative dimension
 $\Delta_0<0$. Then equ.(\ref{eq:Fc})
has to be modified to $c\rightarrow\tilde{c}=c-12\Delta_0$.
This equation is then used to extract the value of $c$ by
numerical finite-size studies or,  as we have done,  using the
Thermodynamic Bethe Ansatz.

\subsection { The representation space of the Virasoro algebra}

In order to construct a Hilbert space,  in which the conformal fields may
act as operators,  we have to go from correlation functions to an operator
formalism. Remember that the charges,  which
via equal time commutators,  genenerate the variations of the
fields under some transformations,  are written as integrals over
space-like surfaces.  Looking at equ.(\ref{eq:QT}), we see that
this integral is in our case $\oint dz T(z) $. This is made explicit
adopting the {\em radial} quantization in the complex plane.
        One considers a Minkowski space
        $\sigma ^0,\sigma ^1 $ with a periodic
         space coordinate $\sigma ^1$. This amounts to
          quantize the theory on a cylinder
       of radius $R$, which then acts as an infra-red cutoff
       for the massless  fields. These separate into
       left- and right-moving fields living on the light-cones
       $\sigma ^0 \pm \sigma ^1$.
       With the map $z=e^{\sigma ^0 +\imath \sigma ^1}$, which we
have already used, the
       cylinder is then mapped onto the complex $z$-plane
       and analogously for right-movers.
The space-like 'equal time' slices are concentric circles around the origin,
whereas the 'time' direction is radially outward, $t=-\infty$
corresponding to $z=0$ and $t=+\infty$ to the point $z=\infty$ of
the complex plane.
Time translations
       $ \sigma^0\rightarrow \sigma^0 + \tau $ on the cylinder
       are mapped into dilatations $ z \rightarrow ze^\tau$
( and $\rev z\rightarrow \rev z e^\tau$ )
       in the plane. Therefore, what we call Hamiltonian on the cylinder,
is the generator of dilatations (= scale transformations) in $z$ and $\rev z$.

Remembering that Euclidean correlation functions correspond to
time-ordered Green-functions, we can consider operators
$\hat{\phi}(z)$ defined such that
\beq
    <\phi_1(z_1)\ldots\phi_n(z_n)>=
   <0|{\cal R}\left(\hat{\phi}_1(z_1)\ldots\hat{\phi}_n(z_n)\right)|0>,
\eeq
where we introduced the {\em radial} ordering operation ${\cal R}$:
\beq
      {\cal R}(\hat\phi_ 1(z_1)\hat\phi_ 2(z_2))=
       \left\{  \begin{array}{ll}
               \hat{\phi}_1(z_1)\hat\phi_2(z_2) & |z_1|>|z_2| \\
               \hat{\phi}_2(z_2)\hat\phi_1(z_1) & |z_2|>|z_1| .
                \end{array}
      \right.
\eeq
With this understanding, we can write equ.(\ref{eq:wardT}) as
\beq
             \delta\hat{\phi}(z,\rev z)   =
       \frac{1}{2 \pi \imath}
      \left( \oint_{|w|>|z|}-\oint_{|w|<|z|} \right) dw\,\epsilon(w)
     R(\hat{T}(w)\hat{\phi}(z,\rev z)           ,
\label{eq:delfi}
\eeq
the difference of the two line integrals giving the integral around
the point $z$ \footnote{ This is what we mean by the equal-time
                        commutator $[\int dx A(x),B(y)]$. For euclidean
                        Green functions, the {\em time} evolution converges
                        only for $\Delta\tau>0$ in $exp(-\Delta\tau\hat H)$.
                        Therefore the ${\cal R}$-operation does give sense to
                        equ.(\ref{eq:delfi}).}.
This is now  an operator statement.

For the particular case of $T(z)$ itself, we get
\bdm
  \delta \hat{T}(z) = \frac{1}{12} c\epsilon^{'''}+
                    (2\epsilon\pri(z)+\epsilon(z)d/dz)\hat{T}(z)
\edm
\beq
         =\left \{ \oint_{|w|>|z|}-\oint_{|z|>|w|} \right \}
          \epsilon(w) R \left ( \hat{T}(z)\hat{T}(w) \right).
\label{eq:deltat1}
\eeq
We now expand $\hat{T}(z)$ and $\hat{\rev T}(\rev z)$ in a Laurent series:
\beq
\hfill          \hat {T}(z)=\sum_{n=-\infty}^{+\infty}
             \frac{L_n}{z^{n+2}},\hspace{3 em}
          \hat{\rev T}(\rev z)=\sum_{n=-\infty}^{+\infty}
                     \frac{\rev L_n}{\rev z^{n+2}},\hfill
\label{eq:Laurent}
\eeq
where the factor $z^{-2}$ has been introduced so that the
operators $L_n$ have dimension $n$ \footnote{ We may expand
         be around any point, not necessarily around the origin
         as done here.}.
If this expansion, together with a similar one for
$\epsilon(z)$,  is now inserted into equ.(\ref{eq:deltat1}),
we obtain the infinite-dimensional Virasoro algebras:
\bdm
     [L_n,L_m]=(n-m)L_{n+m} +\frac{1}{12}cn(n^2-1)\delta _{n+m,0}
\edm
\bdm
     [\rev L_n,\rev L_m]=(n-m)\rev L_{n+m}
              +\frac{1}{12}cn(n^2-1)\delta _{n+m,0}
\edm
\beq
     [L_n,\rev L_m]=0.
\eeq

The space in which the representation of the Virasoro algebra acts
is then constructed as follows. We want to build up a Hilbert space
in which the primary field $\phi _h(z,\rev z) $ with
conformal weight$(h,\rev h)$ may act.
Define the vector
\beq
    |h>\equiv \lim_{z\rightarrow 0} \phi_h (z,\rev z) |0>.
\eeq
{}From equ.(\ref{eq:ward2}) it follows, that
\beq
     [L_n,\phi(z)]=
     \oint \frac{dw}{2\pi\imath}w^{n+1}T(w)\phi(z)=
     h(n+1)z^n\phi(z)+z^{n+1}\partial\phi(z),
\eeq
consequently $ [L_0,\phi(0)]=h\phi(0)$ and
$ [L_n,\phi(0)]=0,n>0$.
The state $|h>$ therefore satisfies
\beq
    L_n |h>=\rev L_n |h> = 0\;\;for\;\; n>0
\eeq
\bdm
      L_0|h>=h|h>,\;\;
           \rev L_0 |h>=\rev h |h>.
\edm
{}From the commutation relations of the $L_n$'s, we see that $L_n$
decreases the dimension by $n$ units, so that the above
 equations guarantee,
that the spectrum of dimensions is bounded from below.

  Notice that $L_{-n},\rev L_{-n}$ for $n>0$ create new states and therefore
act like creation operators,
whereas $L_{n},\rev L_{n}$ are destruction operators. The space
is therefore built up by all vectors of the form
\beq
      L_{-n_1}\ldots L_{-n_N} \rev L_{-m_1}\ldots \rev L_{-m_M}|h>\;\;
       with\;\; n_j,m_k\,>0.
\eeq

These vectors are also generated acting with {\em descendant} fields
on the vacuum. These fields
are the regular terms in equ.(\ref{eq:ward2})
 as $z\rightarrow w$ in the following equation,
which we now write out explicitly :
\bdm
   T(z)\phi(w,\rev w) \equiv \sum_{n \geq 0}(z-w)^{n-2}
         \tilde{ L}_{-n}\phi(w,\rev w)
\edm
\beq
    =\frac{1}{(z-w)^2}\tilde L _0\phi+\frac{1}{z-w}\tilde L _{-1}\phi
     +\tilde L _{-2}\phi +(z-w)\tilde L _{-3}\phi+\ldots.
\eeq
 The descendant fields are
\beq
   \phi^{(-n)}(w,\rev w)=\tilde L _{-n}\phi(w,\rev w)
   =\oint _w \frac{dz}{2 \pi \imath} \frac{1}{(z-w)^{n-1}}
                                     T(z)\phi(w,\rev w).
\label{eq:desc}
\eeq
Comparing with the l.h.s. of equ.(\ref{eq:ward2}), we note that
$ \phi^{(0)}=\tilde L _0\phi=h\phi$ and
$\phi^{(-1)}=\tilde L _{-1}\phi=\partial_z\phi$ \footnote{
         This fact allows to write down differential
          equations satisfied by correlation functions.}.
The most important
descendant field - and therefore not a primary field ! -
is the energy-momentum tensor, which is a level
two descendant of the identity:
\beq
   (\tilde L_{-2}\,I)(w)=\oint\frac{dz}{2 \pi \imath}
              \frac{1}{z-w}T(z)\,I=T(w).
\eeq

{}From equ.(\ref{eq:desc}) and the definition of $L_n$ as
\bdm
     L_n\phi(0)=\oint\frac{dz}{2\pi\imath}z^{n+1}T(z)\phi(0),
\edm
we see that the representation space
is generated by the descendant fields as :
\beq
    L_{-n}|h>=L_{-n}(\phi(0)|0>)=(\tilde L _{-n}\phi)|0>=
     \phi^{(-n)}(0)|0>.
\eeq

This representation is irreducible, unless  there exists a
{\em null} vector
 \bdm
      |\chi _{h+N}>=\sum a_{k\ldots}L_{k_1}\ldots L_{k_m}|h>
\edm
with $\sum k_j=N$,
  satisfying
\beq
   L_n |\chi_{h+N}> =0,n>0 \; \hspace{5em}
   L_0 |\chi_{h+N}>=(h+N)|\chi_{h+N}>
\label{eq:nulvec}
\eeq
for some positive integer N.

For  example for $N=2$, which is the first non-trivial case, these conditions
are
\beq
        L_{+1}(L_{-2}+aL_{-1}^2)\phi(z,\rev z)=0
\label{eq:level2}
\eeq
\bdm
        L_{+2}(L_{-2}+aL_{-1}^2)\phi(z,\rev z)=0
\edm
and conditions for $n\geq 3$ follow automatically from the Virasoro
algebra. Now we move the {\em destruction} operators $L_{+1},L_{+2}$
to the right till they give zero due to equ.(\ref{eq:nulvec}). The
resulting two equations are
\beq
    a=-\frac{3}{2(2h+1)}, \hspace{5em} c=\frac{2h(5-8h)}{2h+1}.
\eeq
We will use analogous equations for level 3 degeneracy in order
to obtain a first nontrivial conservation law in the next chapter.

Since the subspace generated by applying
$L_n,\rev L_m$'s  with $n,m<0$ is invariant, it has to be factored
out in order to get an irreducible representation.
This factor space is called a {\em degenerate }
irreducible representation space (or modul ) generated by the
degenerate field $\phi(z)$ and $N$ is it's level.

These degenerate fields are very important, because for them
the operator algebra \footnote{This equation is written
                              in an abbreviated notation only.}
\beq
  \Phi_l\Phi_k=\sum_n c_{lkn}\Phi_n
\eeq
closes with a finite number of terms. The corresponding central charge
can be labeled by two positive integers $p,p\pri$ with no
common divisor:
\beq
   c(p,p\pri)=1-\frac{6(p-p\pri)}{pp\pri}.
\label{eq:nun}
\eeq
For each value of $c(p,p\pri)$ there are $(p-1)(p\pri-1)/2$ primary fields with
dimensions
\beq
      h_{r,s}=h_{p\pri-r,p-s}=
             \frac{(rp-sp\pri)^2-(p-p\pri)^2}{4pp\pri)}
\label{eq:hpp}
\eeq
\bdm    1\leq r\leq p\pri -1,\hspace{4em}1\leq s\leq p-1.
\edm
For $p\pri-p\geq 2$, there will exist negative weights corresponding to
growing correlation function in non-unitary theories.
For $c<1$ all  unitary theories are special cases of the above
with $p\pri-p=1$ or :
\beq
           c(m)=1-\frac{6}{m(m+1)},\,\;m=3,4,\ldots
\label{eq:deg}
\eeq
For each value of $c(m)$ there are again $m(m-1)/2$ allowed values of $h$:
\beq
      h_{r,s}(m)=h_{m-r,m+1-s}=\frac{[(m+1)r-ms]^2-1}{4m(m+1)}
\label{eq:wei}
\eeq
with the integers $r,s$ satisfying $1\leq r \leq m-1,1\leq s \leq m$.

We also know that the representation with highest weight $h_{rs}$ is
degenerate at level $rs$.
Besides this the correlation functions of these fields satisfy
linear differential equations of order $rs$.

We may now clarify the shift $c\rightarrow\tilde c$ for
non-unitary models. Recalling equ.(\ref{eq:TLL}) of section~\ref{s3},
 the quantum hamiltonian $\hat H$ is given by:
\bdm
   \hat H=\frac{1}{2\pi}\int_{0}^{R}\;T_{uu}(v)dv=
          \frac{1}{2\pi}\int_{0}^{R} (T(v)+\rev T(v))dv=
\edm
\beq
        \frac{2\pi}{R}(L_0+\rev L_0) - \frac{\pi c}{6R}.
\eeq
As advertised, we see that time evolution on the cylinder, corresponds
to dilatations on the plane. From equ.(\ref{eq:TLL}), it follows that
states with higher eigenvalues of $L_0$ and $\rev L_0$, contribute
 exponentially smaller corrections to $Z(L,R)$. However in non-unitary
theories some  operators may have negative dimensions. Suppose there
is one such operator with dimension $\Delta_0<0$. In this case a
factor $exp(-2\pi\Delta_0/R$ has to be kept in $Z(L,R)$, effectively
shifting $c$ to $\tilde{c}=c-12\Delta_0$.

\subsection{ Characters of the Virasoro algebra}

For Zamolochikov's counting argument, which permits a partial
identification of the infinite number of conservation laws
surviving perturbations breaking conformal invariance,  we need to
determine the dimensional decomposition of the representations of the
Virasoro algebras. In the representation space of highest weight
$h$ the {\em characters}  $\chi_h(q)$
are the generating functions for the number of linearly independent
vectors at level $n$, therefore having eigenvalues $h+n$ of $L_0$.
For $q,|q|<1$ we define
\beq
 \chi _h(q) \equiv q^{-c/24}\,Tr_h\,q^{L_0}=
q^{h-c/24}\sum_{n=0}^{\infty}\,d_h(n)\,q^n.
\eeq
Here $d_h(n)$ counts the degeneracy of the states in the representation
at level $n$.
An analogous definition holds for the right Virasoro algebra.

If there are no null states in the representation of weight $h$, the states
at level $n$ are of the form
\beq
   L_{-n_1}L_{-n_2}\ldots L_{-n_k}|h>\hspace{4em} \sum_{i}^k\,n_i=n.
 \eeq
In this case $d_h(n)$ equals $p(n)$ - the number of
partitions of the integer $n$. Euler's generating function for these
partions gives
\beq
       \sum_{n=0}^{\infty}\,q^np(n)=\prod_{n=1}^{\infty}(1-q^n)^{-1}
        \equiv \frac{1}{q^{-1/24}\,\eta(q)}
\label{eq:eta}
\eeq
and
\beq
       \chi_h(q)=q^{h+(1-c)/24}\,\eta (q)^{-1}.
\eeq
Here we have introduced Dedekind's $\eta$-function, central in the
study of elliptic functions. The $c$-dependent factor has introduced, so that
$\chi_{rs}(q)$ has nice {\em modular} transformation properties.

Let us compute the characters for the degenerate, unitary series with
$c<1$. The allowed weights are  given by equ.(\ref{eq:wei}) and are
degenerate at level $rs$.
Thus in counting the states we have to subtract the null state at
level $rs$ and all its descendants, getting
\beq
     \chi_{r\,s}(q)=q^{(1-c)/24}\eta(q)^{-1}( q^{h_{r,s}} -
                               q^{h_{r,s}+rs} + \ldots ).
\eeq
     But the null state has weight $h_{r,s}+rs=h_{r,-s}=h_{m+r,m+1-s}$
and therefore we have to subtract in turn it (and all its descendants)
at level $(m+r)(m+1-s)$. Correcting the above formula for $\chi_{r\,s}$,
we get
\bdm
    \chi_{r\,s}(q)=q^{(1-c)/24}\eta(q)^{-1}(q^{h_{r,s}}-q^{h_{r,-s}}(
              1-q^{(m+r)(m+1-s)}+\ldots))
\edm
\beq
        =q^{(1-c)/24}\eta(q)^{-1}(q^{h_{r,s}}-q^{h_{r,-s}}+q^{h_{2m+r,s}}
               -\ldots).
\eeq
Repeating this process yields finally the  correct expression
for the character:
\beq
    \chi_{r\,s}(q)=q^{(1-c)/24}\eta(q)^{-1}
         \sum_{k=-\infty}^{\infty}\,(q^{h_{2mk+r,s}}-q^{h_{2mk+r,-s}}).
\label{eq:degch}
\eeq
\sect{ Deformations of Conformal Invariant Field Theories }{s5}

In this section we will study relevant perturbations of Conformal Invariant
Field \hspace{0.2cm}Theories ( CFT ), i.e. perturbations which drive
the system
away from
its UV critical unstable fixed point.  An example would be the perturbation
of a statistical mechanical system at its critical point by a
magnetic field, the
raising of the temperature off $T=T_c$ etc. The system will then flow away
from its UV fixed point and may either end up at another critical conformally
invariant fixed point or develop a finite correlation length, i.e.  the
theory becomes massive.

In a CFT conservation laws are trivially satisfied by fields, which depend
only on either of the light cone coordinates  $z$ or $\rev{z}$,
for example the energy-momentum tensor and it's regularized powers
( in most of the models, we are going to look at, there are also
other conserved currents ).
If an infinite number of local integrals of motion survives the
perturbation of the CFT, then we know that the massive theory is
described by factorized $S$-matrices, of which some examples have
been studied in chapter \ref{s3}.
There we made some assumptions, which resulted
in the existence of an infinite number of conserved charges $P_s$, where
$s$ belonged to a certain subset of the natural numbers \footnote {
       Recall that $P_{-s}$ is related to $P_{+s}$ by parity, which
       we always assume to be a symmetry, allowing us to restrict ourselves
       to $s>0$. }.
Suppose we perturb the critical action $S^*$ by a relevant scalar
operator:
\beq
     S = S^* - \lambda \int \phi(z,\rev{z})\,d^2z,
\eeq
where the weight of $\phi$ is $(h,h)$, so that the
dimension of $\lambda$ is $(1-h,1-h)$. For this to be a relevant
perturbation, we need $y=2(1-h)>0$ or $ h\,<\,1 $.
Charges, whose conservation is preserved by the perturbation, now become
\beq
     P_s = \oint \,[ T_{s+1} \, dz \; + \Theta_{s-1} \,d\rev{z}],
\eeq
where $T_s$ and $\Theta_s$ are local fields of spin $s$, satisfying the
continuity equation
\beq
           \partial_{\rev{z}} \,T_{s+1} =
           \partial_{z} \,\Theta_{s-1}  .
\label{eq:cont1}
\eeq

If $S^*$ described a free field theory, such a perturbation would
be called su\-per-re\-nor\-mali\-zable and adding the finite number of
terms with dimensions smaller or equal than that of $\phi$ one obtains
a finite theory without UV-divergencies. In particular this
doesn't change the structure of the UV fixed point and the fields
continue to have the same dimensions at short distances, although
the large distance behaviour of the perturbed theory is very much
different. We will assume that this is also true for our case, where
we perturb around a theory, which is not free.

Thus $S^*$ describes
an UV-finite ( renormalized ) theory, which contains $\phi$ as one
of its operators and we assume,
that we know all its correlation
functions ( as well as those of all other operators ). Our primary aim
is to check whether there exist currents $J(z)$,
whose conservation survives the perturbation,
i.e. whether it satisfies equ.(\ref{eq:cont1}).

The perturbed correlation functions  of a particular operator $J(z)$
 are now given by
\beq
   <\,J(z,\rev {z})\ldots>=
   <\,J(z)\ldots>_{S^*}
+ \lambda \int \, d^2z_1 \, <\,J(z) \phi(z_1,\rev{z}_1)\ldots>_{S^*}
+ {\cal O}(\lambda^2).
\label{eq:per1}
\eeq

If this integral were finite, it would be independent of $\rev{z}$.
Therefore any $\rev{z}$-dependence can come only from possible
singular points $z \rightarrow z_1$. In their neighborhood, we
can use the short distance expansion ( SDE ) :
\beq
  J(z) \,  \phi(z_j,\rev{z}_j) =
\sum_k \; \frac{a_k}{ [z-z_j]^{\Delta_J+\Delta-\Delta_k} }
               \phi_k(z_j,\rev{z}_j),
\label{eq:OPEJ}
\eeq
where $\Delta=2h$ and $\Delta_J$ and $\Delta_k$ are the scaling dimensions
of $J$ and $\phi_k$. Since only non-integrable
singularities will contribute,
this requires $\Delta_J+\Delta-\Delta_k \geq 2 $. In a unitary theory
all dimensions are $>0$ and therefore only a finite number of
 operators $\phi_k$ will
contribute to first order in $\lambda$.

Let us look at the specific example of the energy-momentum tensor. In this
case the SDE is
\beq
    T(z)\phi(z_1, \rev{z}_1) =
                     \frac{h}{(z-z_1)^2}\phi(z_1,\rev{z}_1)+
                \frac{1}{z-z_1}\partial_1\phi(z_1,\rev{z}_1).
\label{eq:OPET}
\eeq
Since we only want to check whether the current $T(z,\rev z)$
 is still conserved
and to avoid bothering with infra-red singularities, let us only calculate
the $\partial_{\rev z}$ derivative of $T(z,\rev z)$.
For this purpose we use the equation
\beq
     \partial_{\rev{z}}(z-\xi)^{-m-1} =
      2\pi\imath  \frac{(-1)^m}{m!} \partial^{m}_{z}\delta^{(2)}(z-\xi).
\eeq
This can easily be proved, either remembering that the logarithm is the
2-dimensional Green function $\triangle \ln r = -2\pi\delta^{(2)}(\vec{r})$,
i.e.
\bdm
     \partial_z\partial_{\rev z}\ln[(z-z\pri)(\rev z-\rev z\pri)]=
     -2\pi\imath\delta^{(2)}(z-z\pri)
\edm
and taking derivatives thereof, or proceeding as follows.
Let us regulate the UV divergence by a cut-off $a$ ,
inserting the step function $H(x)$, which vanishes for $x<0$. Then
\bdm
   \int d^2z\, \partial_{\rev{z}}
     \frac{H[z\rev{z}-a^2]}{ z^{m+1} } f(z,\rev{z})=
\edm
\beq
    \int d^2z \frac{ z\delta (z\rev{z}-a^2) }{z^{m+1} }f(z,\rev z) =
       \imath\int d\varphi \frac{  f(ae^{\imath\varphi},ae^{-\imath\varphi})  }
                          { a^m\,e^{\imath m\varphi} }=
\eeq
\bdm
   2\pi\imath \frac{f^{(m)}_z(0,0)}{m!}=2\pi\imath \frac{(-1)^m}{m!}
                       \int d^2z\,\partial^m_{z}\,\delta^{(2)}(z)
                        f(z,\rev z).
\edm
Hence $T(z,\rev z)$ satisfies
\bdm
   \partial_{\rev z} \int d^2 z_1 <T(z)\varphi(z,\rev z)\ldots>=
   \int d^2 z_1\partial_{\rev z} <\left(
                     \frac{h}{(z-z_1)^2}
\phi(z_1,\rev{z}_1)+ \frac{1}{z-z_1}\partial_1\phi(z_1,\rev{z}_1)\right)
             \ldots>=
\edm
\beq
         \int d^2z_1  \left( h2\pi\partial_z\delta^{(2)}(z-z_1)+
                2\pi\delta^{(2)}(z-z_1)\partial_1\right)
      <\varphi(z_1,\rev z_1)\ldots>.
\eeq

We therefore immediately get the conservation law for the energy-momentum
tensor - as expected, since the energy-momentum tensor must remain
conserved - :
\beq
    \partial_{\rev{z}} T + \partial_{z}\Theta=0,
\label{eq:dzbarT}
\eeq
where
\beq
   \Theta = \pi \lambda(1-h)\phi(z, \rev{z}).
\eeq
We see, that the term in the SDE equ.(\ref{eq:OPET}),
relevant for the conservation of the current $J(z,\rev z)$ is picked out by :
\beq
  \partial_{\rev{z}} J(z,\rev{z})
          = \lambda \pi \oint_{C_z} \frac{dz_1}{2\pi \imath}
                  J(z) \phi(z_1,\rev{z}_1).
\label{eq:Jcons}
\eeq
Remembering the discussion on radial quantization leading to
equ.(\ref{eq:delfi}), we realize that the r.h.s. of this equation
is a commutator and we get the following suggestive form :
\beq
     \partial_{\rev z} J(z,\rev z)=
      [J(z,\rev z),H_{int}(\rev z)],
\eeq
where
\bdm
     H_{int}(\rev z)=\lambda\int dz_1 \phi(z_1,\rev z).
\edm
It remains to be seen, if the r.h.s. of equ.(\ref{eq:Jcons})
can be expressed as
$\partial_z$ of some operator. If yes, the conservation law will
continue to hold to first order in $\lambda$.

To provide an example of how this can be checked, let us take the
energy-momentum tensor and it's powers.
Define $\Lambda$  to be the irreducible  Virasoro
modul with highest weight $h=0$,  to  which the energy-momentum tensor
belongs. Introduce Virasoro generators as in equ.(\ref{eq:Laurent}), but
via an expansion around an arbitrary point $\zeta$ :
\beq
    T(z)=\sum_{n=-\infty}^{+\infty} \frac{L_n}{(\zeta-z)^{n+2}}.
\eeq
Then, using  the commutation relation
\beq
    [L_{-1},\phi(\zeta,\rev{\zeta})] =
             \partial_{\zeta}\phi(\zeta,\rev{\zeta}),
\eeq
it immediately follows for the operator $\partial_{\rev z}$ in
 equ.(\ref{eq:Jcons}) - with $J(z)$ replaced by $T(z)$ -
\beq
      \partial_{\rev z}L_{-1}\Lambda=
      L_{-1}\partial_{\rev z}\Lambda.
\eeq

Now we streamline our algebra  following reference~\cite{ZAMOT}.
Introduce a set of operators $D_n,\,\,n=0,\pm 1,\pm 2,\ldots$ as
\beq
   D_n\Lambda(z,\rev z)=
    \oint_z \frac{d\zeta}{2\pi\imath} \phi(\zeta,\rev z)
            (\zeta-z)^n \Lambda(z),
\eeq
i.e. $D_n$ projects out the term proportional to $(\zeta-z)^{n+1}$
in the SDE of $\phi$ and $\Lambda$.
The following equations show, why the $D_n$ are of good use to compute
$\partial_{\rev z}$ :
\beq
          \partial_{\rev z}=D_0.
\eeq
Also
\bdm
    D_{-n-1}I=\oint\frac{d\zeta}{2\pi\imath}(\zeta-z)^{-n-1}
              \phi(\zeta,\rev z),\,\,n\geq 0,
\edm
and using the residue theorem for the (n+1)-th order pole :
\beq
     D_{-n-1}I=\frac{1}{n!}\partial_z^n\phi(z,\rev z).
\eeq
Since $T(z)$ is generated from the unity $I$ by applying operators
$L_{-n}$, we need the commutation relations between $L_{-n}$ and
$D_m$. First remember
\beq
  [L_{-n},\phi(\zeta,\rev{\zeta})]=
\{ (\zeta-z)^{n+1}\partial_{\zeta}+(n+1)h(\zeta-z)^n \}
                              \phi(\zeta,\rev{\zeta}).
\eeq
Now compute the commutator of $[L_n,D_m]$ :
\bdm
 L_nD_m\Lambda(z,\rev z)=
      \oint_z\frac{d\zeta}{2\pi\imath}L_n\phi(\zeta,\rev z) =
\edm
\bdm
              (\zeta-z)^m \Lambda(z)=
   \oint_z \frac{d\zeta}{2\pi\imath}
         \left( [L_n,\phi(\zeta,\rev z)] +\phi(\zeta,\rev z)L_n \right)
         (\zeta-z)^m\Lambda(z).
\edm
Therfore
\bdm
   [L_n,D_m]\Lambda(z,\rev z)=
    \oint_z\frac{d\zeta}{2\pi\imath}[L_n,\phi(\zeta,\rev z)]
             (\zeta-z)^m\Lambda(z)=
\edm
\bdm
   \oint\frac{d\zeta}{2\pi\imath} \left(
         (\zeta-z)^{n+m+1}\partial_{\zeta} +(n+1)h(\zeta-z)^{n+m} \right)
         \phi(\zeta,\rev z)\Lambda(z)=
\edm
\bdm
   \{-\left( (1-h)(n+1)+m\right)\}
   \oint_z\frac{d\zeta}{2\pi\imath}\phi(\zeta,\rev z)(\zeta-z)^{n+m}\Lambda(z).
\edm
Hence we get the commutator
\beq
 [L_n,D_m]=-\{(1-h)(n+1)+m\}D_{n+m}.
\eeq
A trivial application is
\beq
   \partial_{\rev z}T(z,\rev z)=\lambda D_0 L_{-2}I=
   \lambda(h-1)D_{-2}I=\lambda(h-1)L_{-1}\phi(z,\rev z)
\eeq
reproducing equ.(\ref{eq:dzbarT}).

A less non-trivial calculation is to check the conservation of higher powers
of the momentum. Let us define a regularized square of $T$
$T_4(z)= :T^2(z):$ as
\beq
 T_4(z)\equiv (L_{-2}L_{-2}I)(z)=\oint_zd\zeta(\zeta-z)^{-1}T(\zeta)T(z).
\eeq
Now check it's conservation :
\bdm
  \partial_{\rev z}T_4=\lambda D_0 L_{-2} L_{-2}I=
   \lambda (h-1)(D_{-2}L_{-2}+L_{-2}D_{-2} )I=
\edm
\beq
   =\lambda(h-1)(2L_{-2}L_{-1}+\frac{h-3}{6}L_{-1}^3)\phi.
\label{eq:consT4}
\eeq
Therefore, due to the first term above, the r.h.s. is in general not a
derivative of $z$. This is to be expected, since the existence or not of this
conservation law is a dynamical question.
Let us take as perturbation one of the fields $\phi_{1,3}$ of the unitary
models with $c<1$ as an example. It is degenerate at level 3
 and repeating the steps that led to equ.(\ref{eq:level2}) for the present
case, we would get the following null-vector equation:
\beq
 \left (L_{-3}-\frac{2}{h+2)}L_{-1}L_{-2}
  +\frac{1}{(h+1)(h+2)}L_{-1}^3 \right )\phi_{1,3}(z)=0.
\eeq
Hence the term containing $L_{-2}$ in equ.(\ref{eq:consT4}) can be eliminated
in favor of the derivative $L_{-1}$~: $L_{-2}L_{-1}=L_{-1}L_{-2}-L_{-3}$ and
using for $L_{-3}$ the null-vector equation, we get
\beq
  \partial_{\rev z} T_4(z,\rev z)=\partial_z \Theta_2(z,\rev z),
\eeq
with
\bdm
 \Theta_2=\lambda\frac{h-1}{h+2}\left(
   2hL_{-2}+\frac{(h-2)(h-1)(h+3)}{6(h+1)}L_{-1}^3\right)\phi_{1,3}.
\edm

Finally let us check  that, similarly to the first order perturbation,
the perturbation expansion to n-th order, which usually
contains an {\em infinite } number of terms, is here also
drastically truncated. The n-th order term will have the form
\beq
  \lambda^n \int \, d^2z_1\ldots d^2z_n
   <\,J(z) \prod_{j=1}^n \phi(z_j,\rev{z}_j)\ldots>_{S^*}
\label{eq:pern}
\eeq
We easily see , that the condition for non-integrable singularities is
now $\Delta_J+n\Delta-\Delta_k > 2n $ or
\makebox {$\Delta_J-(2-\Delta)n-\Delta_k \geq 0 $ }.
Since $\Delta < 2$, this condition will eventually be violated
and the perturbation expansion  has to stop with a {\em finite} number of
terms.

\subsection{ Counting Arguments }

In principle we may now take specific models and start looking for
surviving conservation laws. It turns out \cite{ZAMOT}, that at the
expense of some formalism this job can be significantly simplified.
Let us then present, what is called Zamoldochikov's {\em counting
argument}. It will enable us to find out which of the $P_s$ are still
conserved, at least for small $s$, without having to compute
explicitly the term $ \Theta_s $ in equ.(\ref{eq:cont1}). Once we know
the conservation laws, we may link up with chapter \ref{s1} and lift a
candidate $S$-matrix for the model in question.

Let us talk about the energy-momentum tensor $T$, since $T$
 and
its regularized powers will provide the conservation of the
momentum and it's powers. The same type of reasoning also
applies to other conserved currents.

Recall that $\Lambda$ was defined to be the irreducible  Virasoro
modul with highest weight $h=0$,  to  which the energy-momentum tensor
belongs. That is, $\Lambda$ is the infinite-dimensional space spanned by
all the fields of the form $ L_{-n_1}L_{-n_2}\ldots L_{-n_k}\, I $, where
$n_i$ are positive integers or zero. $\Lambda$ may be decomposed as
\beq
    \Lambda=\bigoplus_{s=0}^{\infty}\;\Lambda _{s},
\eeq
where the fields belonging to $\Lambda _{s}$ satisfy
$\sum_{i=1}^k n_i=s$. From the commutation relations for the $L_n$ we
easily see that the $\Lambda _{s}$ are eigenspaces of $L_0$:
\beq
               L_0\Lambda _{s}=s\Lambda _{s}.
\eeq
Thus all fields belonging to $\Lambda _{s}$ have conformal weight
$(s, 0)$ and therefore dimension and spin equal to $s$. Besides this all
these fields depend only on $z$ and are thus analytic, satisfying
$\partial_{\rev z}\Lambda = 0$ : they all give rise to integrals
of motion.
However for our counting argument it is important
to exclude fields, which are derivatives of others. These would lead
 situations like $\partial_z\partial_{\rev z}T=\partial_{\rev z}R$,
which do not correspond to conservation laws at all.
They are contained
in $L_{-1}\Lambda$. Therefore let us define a new space $\hat{\Lambda}$, where
these fields are divided out : $\hat{\Lambda}=\Lambda/\L_{-1}\Lambda$. This
space also has the decomposition
\beq
      \hat{\Lambda}=\bigoplus_{s=0}^{\infty}\, \hat{\Lambda} _{s},
      \hspace{2em}
      L_0\hat{\Lambda} _{s}=s\hat{\Lambda} _{s}.
\eeq
We can now take advantage of character formulas to obtain the dimensions
of $\hat{\Lambda} _{s}$:
\beq
   \sum_{s=0}^{\infty}\, q^s\, dim\hat{\Lambda} _{s}\,=
      (1-q)\chi_0(q)+q,
\label{eq:charl}
\eeq
where $\chi_0(q)$ is the character of $\Lambda$ defined as
\beq
     \chi_0(q)=\sumzi\, q^s\,dim\Lambda _{s}.
\eeq
Equ.(\ref{eq:charl}) can easily be shown as follows.
First note that
\bdm
  dim\left( (L_{-1}\Lambda) _{s}\right)=\left\{
  \begin{array}{ll}
             dim\Lambda _{s-1}\,,\,for &   s > 1  \\
                         0             &   s=1,
  \end{array}
                                        \right.
\edm
since $L_{-1}I=0$.
Therefore  :
\bdm
  \sumzi q^s dim\left( (L_{-1}\Lambda)_{s}\right)=
   \sum_{s=2}^{\infty} q^s dim\Lambda _{s-1}=
    q\,\sum_{s=2}^{\infty}q^{s-1}\, dim\Lambda _{s-1}=
\edm
\bdm
    q\, \sumzi\, q^s\,dim\Lambda _{s}-q=
    q\, \chi_0(q)-q.
\edm
and equ.(\ref{eq:charl}) follows.
If $c$ doesn't belong to the degenerate set given by
equ.(\ref{eq:nun}) or equ.(\ref{eq:deg}),
the dimension of
$\Lambda _s$, according to its definition,  equals $p(s)$ - the number of
partitions of the integer $s$. Equ.(\ref{eq:eta}) gives then
\beq
       \sumzi\, q^sp(s)=\prod_{n=1}^{\infty}(1-q^n)^{-1}
       =\chi_0 (q),
\eeq
so that finally we get \footnote{
          For the degenerate cases, we have to subtract invariant subspaces
          and use equ.(\ref{eq:degch}). Even then equ.(\ref{eq:charl}) is
          still valid for $s<m(m-1)$.}:
\beq
   \sumzi\, q^sdim(\hat{\Lambda} _s)=1+q^2+q^4+2q^6+\ldots.
\eeq

Now we have to ask, which of these conserved currents survive
perturbation ?
Suppose we add a relevant perturbation $\lambda \phi(z, \rev z) $.
$\phi(z, \rev z)$ is the highest weight vector of the
irreducible modul $\Phi\bigotimes\rev\Phi$, generated by
all the fields of the form
$ L_{-n_1}\ldots L_{-n_k}\rev L_{-m_1}\ldots\rev L_{-m_l}\phi $, where
$n_i,m_j$ are positive integers or zero.
As $\Lambda$,  $\Phi$ ( and $\rev\Phi$  )  may be decomposed as
\beq
    \Phi=\bigoplus_{s=0}^{\infty}\;\Phi _{s},\,
    L_0\Phi _s = (\Delta + s)\Phi _s,\,
    \rev {L} _0 \Phi _s=\Delta \Phi _s\,.
\eeq
Now the fields  $T_s^{(\kappa)} \in \hat{\Lambda}_{s}$ do not satisfy
$\partial_{\rev z} T_{s}^{(\kappa)}=0$, but
\beq
    \partial_{\rev z} T_{s}^{(\kappa)}=
    \lambda \Phi_{s-1}^{(\kappa)} + O(\lambda).
\eeq
Here $ \Phi_{s-1}^{(\kappa)}$ are local fields belonging to
$\Phi_{s-1}$ and for simplicity assume, that only the first order
term in $\lambda$ contributes \footnote{See ref.\cite{ZAMOT} for
                               a discussion when this is not true.}.
We use this equation to define the linear operator
\beq
      \partial _{\rev z} \;:\;\hat{\Lambda}_s\;\rightarrow\;\Phi_{s-1}.
\eeq
Let us  consider the space $\hat{\Phi}$, where we factored out the
de\-ri\-va\-tives : $ \hat{\Phi}_s=\Phi _s/L_{-1}\Phi _{s-1} $.
Consider now the
mapping $ {\cal M}_s$ from $\hat{\Lambda}_s \rightarrow \hat{\Phi}_{s-1}$.
${\cal M}_s$ is implemented by $ {\cal M}_s= \Pi _s D_{0, s} $, where
$\Pi _s$ is the projector from $\Phi _s$ to $\hat{\Phi} _s$ and
$D_{0,s} $ is $\partial _{\rev z} $ restricted to $\hat{\Lambda}_s $.
Since in $\hat{\Phi}$ we factored out the derivatives, all the fields
satisfying
\beq
      \partial_{\rev z} T_{s+1} = \partial_z \Theta_{s-1},
\label{eq:cons11}
\eeq
are mapped into the null element of $\hat{\Phi}_{s-1}$, i.e.
all elements $T_{s+1}$ satisfying equ.(\ref{eq:cons11}) belong to the
kernel of ${\cal M}_s$. If $dim\,\hat{\Lambda}_{s+1}\,>\,dim\,\hat{\Phi}_s$,
then $Ker\,{\cal M}_s \neq 0$ and we have conserved charges
surviving the perturbation.

The dimension of $\hat{\Phi}_s$ can be computed as that of $\hat{\Lambda}_s$
and we obtain:
\beq
         \sumzi\, q^{\Delta+s}dim\hat{\Phi}_s\,=\,(1-q)\chi _{\Delta}(q),
\label{eq:charf}
\eeq
where $\chi_{\Delta}(q)$ is the character of  the modul with
highest weight $(h,h)$ \footnote{
                     Notice that the identity is the only primary field
                     satisfying $L_{-1}I=0$, which accounts for the
                     difference between equs.(\ref{eq:charl}) and
                     (\ref{eq:charf}).}.

Let us apply this scheme to specific cases.
\\

{\bf 1) Lee-Yang edge singularity and the associated non-unitary series}

The Lee-Yang singularity describes the critical behavior of an
 Ising model in a purely
 imaginary magnetic field $ih$. Above a critical value $h_c$ the zeroes
of the partition function condense on the imaginary $h$-axis and the
following Ginzburg-Landau lagrangian describes this behavior:
\beq
     {\cal L}= \int d^2x [\frac{1}{2}(\partial_{\mu} \varphi(x))^2
              -\imath (h-h_c) \varphi(x) -\imath \varphi(x)^3].
\eeq
This is clearly a non-unitary theory and at the critical point
$h=h_c$ $\varphi(x)$ is the only relevant operator. In ref.\cite{CIZ}
it was shown that there exists only one model with these properties
and no extra symmetries. With the aid of equs.(\ref{eq:nun}) and
equ.(\ref{eq:hpp}) one easily identifies it to have $p=2$ and
$p\pri=5$ with central charge is $c=c(2,5)=-22/5$. It contains
only two primary fields :
the identity and the field $\varphi=\varphi_{(1,2)}=\varphi_{(1,3)}$
 with weights $(0,0)$ and $(-1/5,-1/5)$
respectively. The negative values reflect lack of unitarity here.

We now check for conservation laws computing the dimensions
 of the spaces $\hat{\Lambda}_s$ and $\hat{\Phi}_s$,
using equ.(\ref{eq:charl}) and equ.(\ref{eq:charf}).
One finds, that the
dimension of $\hat{\Lambda}_s$ exceeds the one of $\hat{\Phi}_s$ by
one unit for
\beq
   s= 1,5,7.
\eeq
These are the first 3 conservation laws found in section \ref{s1}.
This reasoning has been extended\cite{CHIC} to
the whole series of non-unitary models
$p=2,p\pri=2N+3$ with
\beq
      c(2,2N+3)=-\frac{2N(6N+5)}{2N+3}
\eeq
perturbed by the field $\varphi_{(1,3)}$.
\\

{\bf 2) Ising model.}

It's conformal anomaly is
 $c=c(m=3)=1/2$. This model contains three spinless primary fields :
$I=\varphi_{(1,1)},\sigma=\varphi_{(1,2)}$ and $\epsilon=\varphi_{(1,3)}$
with conformal weights $(0,0),(1/16,1/16)$ and $(1/2,1/2)$, which are
identified with the identity operator, the spin density (magnetization) and
the energy density respectively. All the other local fields in this model
are obtained by applyig the left and right Virasoro generators $L_n$ and
$\rev L_n$ with $n\,<\,0$ to these primary fields.

Suppose we perturb this model by a magnetic field, which
breaks the $Z_2$ invariance
and couples to the magnetization $\sigma (x)=\varphi _{(1,2)}(x)$.
The total action is
\beq
           S_{1/2}^{(1,2)} = S^* + h\int\,d^2x\, \sigma(x).
\eeq
The dimensions of the spaces $\hat{\Lambda}_s$ and $\hat{\Phi}_s$ can be
calculated, using equ.(\ref{eq:charl}) and equ.(\ref{eq:charf}). For the
characters, we use equ.(\ref{eq:degch}).
The relevant formulas are
\beq
   \chi_0(q)=\chi_{1,1}(q)=\frac{1}{2}\left\{
       \prod_{n=1}^{\infty} (1+q^{n+1/2})+
       \prod_{n=1}^{\infty} (1-q^{n+1/2})\right\},
\eeq
\beq
   \chi_{1/16}(q)=\chi_{1,2}(q)=q^{1/16}\prod_{n=0}^{\infty}
                   (1-q^{2n+1})^{-1}=
                  q^{1/16}\prod_{n=1}^{\infty}
                   (1+q^{n}).
\eeq
Computing now the dimensions, we check that the
dimension of $\hat{\Lambda}_s$ exceeds the one of $\hat{\Phi}_s$ by
one unit for
\beq
    s=1,7,11,13,19.
\eeq
For larger values of $s$ the dimension of $\hat{\Phi}_s$ is greater or
equal
and nothing can be concluded. However we obtained five nontrivial conservation
 laws for the Ising model in a magnetic field and one conjectures the
infinite set, corresponding to all integers $s$, relatively prime to $30$
as found in section \ref{s1}.

\vspace{0.7cm}

{\bf 3) Z(N)-models.}

Using again the by now familiar scheme let usconsider the Z(3) model
perturbed by the field $\varphi_{(1,2)}$, which is the most relevant thermal
operator. We get conservation laws for $s=1,2,4,5,7,8$, which are the first
six integers of the infinite series $s \neq 0,mod(3)$.

This is a particular case of the following general results obtainable
using the machinery developped in the lectures.

The UV limiting CFT of the $Z(N)$ models exhibit
 primary fields of integer spin
$s=3,4,\ldots,N$, which depend
only on the variable $z$ - and analogously for $\rev z$. These are therefore
conserved currents and the Virasoso algebra is enlarged to a $W(N)$
algebra, containing the energy-momentum tensor and those currents \cite{FL}.
This additional symmetry is also suggested by the {\em coset
construction} \cite{GKO}, from which the $Z(N)$ models may be obtained,
using the $A_{N-1}$ Lie algebra.
The central charge of the $W(N)$-models depends on two integers:
\beq
    c(p,N)=(N-1)\left (1-\frac{N(N+1)}{p(p+1)}\right ).
\eeq

For $p=N+1$ we obtain our $Z(N)$ series.
The generalization of the $\varphi_{(1,2)}\in Z(3)$ is a primary
field of weight $(\frac{2}{h^{{\cal A}}+2},\frac{2}{h^{{\cal A}}+2})$,
 which preserves the
global symmetry and is the most relevant thermal operator.
Here $h^{{\cal A}}=N$ is the so-called dual Coxeter number of the
Lie group $A_{N-1}$. Perturbing
the UV limit with this operator, we obtain our massive $Z(N)$ models with
conservation laws $P_s$, where $s$ takes the {\em exponents}
of $A_{N-1}$ ( which are all integers $1,2,\ldots,N$ ),
repeated modulo the Coxeter number $h^{{\cal G}}$, i.e. modulo $N$.

%
\vspace{1cm}

{\bf ACKNOWLEDGMENTS}
 It is my pleasure to thank Drs. M.Martins,
 F.C.Alcaraz and J-B.Zuber for
helpful discussions and the organizers Drs. O.Eboli and M.O.C. Gomes for
giving me the opportunity to lecture at this summer school. I also thank
the students, especially S.Dahmen, S.Rocha Neto and A.Malvezzi,
 for their feedback, which made me learn a lot and thus the
greatest beneficiary of these lecture notes.

\end{document}